\documentclass[11pt]{article}

\ifdefined\pdfoutput
  \pdfoutput=1
\fi

\newlength{\abstractwidth}
\usepackage{amsmath}
\usepackage{amsfonts}
\usepackage{amssymb}
\usepackage{bm}
\usepackage{appendix}
\usepackage{verbatim}

\numberwithin{equation}{section}

\usepackage{graphicx}
\usepackage{xcolor}
\usepackage[export]{adjustbox}
\usepackage{caption}
\usepackage{subcaption}

\usepackage{tikz}
\usetikzlibrary{
  decorations.pathmorphing,
  decorations.pathreplacing,
  arrows.meta,
  positioning
}

\usepackage{hyperref}

\definecolor{darkred}{rgb}{0.8,0.1,0.1}
\hypersetup{
  colorlinks=true,
  linkcolor=darkred,
  citecolor=blue,
  urlcolor=blue,
  linktoc=page
}

\graphicspath{{Figures/}}

\newcommand{\be}{\begin{equation}}
\newcommand{\ee}{\end{equation}}

\def\bb{{\bf b}}

\def\bz{{\bf z}}

\def\bsa{{\boldsymbol{a}}}
\def\bsb{{\boldsymbol{b}}}
\def\bsx{{\boldsymbol{x}}}
\def\bsy{{\boldsymbol{y}}}

\def\bsk{{\boldsymbol{k}}}
\def\bsq{{\boldsymbol{q}}}

\def\bsX{{\boldsymbol{X}}}

\def\L{{\rm L}}
\def\H{{\rm H}}

\def\p{\partial}

\def\eps{\epsilon}
\def\f{\varphi}

\def\m{\mu}
\def\n{\nu}

\def\({\left(}
\def\){\right)}
\def\[{\left[}
\def\]{\right]}

\def\no{\nonumber}

\newcommand{\Isoft}{\mathcal{I}_{\rm soft}}
\newcommand{\IRegge}{\mathcal{I}_{\rm Regge}}

\newcommand{\kv}{\boldsymbol{k}}
\newcommand{\qv}{\boldsymbol{q}_2}
\newcommand{\bv}{\boldsymbol{b}}
\newcommand{\ths}{\boldsymbol{\theta}_s}
\newcommand{\tht}{\boldsymbol{\theta}}

\newcommand{\Th}{\Theta}
\newcommand{\bth}{\bm{\theta}}
\newcommand{\bTh}{\bm{\Theta}}

\newcommand{\fp}{\varphi}

\begin{document}
\thispagestyle{empty}

\vskip 0.3in

\begin{center}
{\Large \bf  Tidal effects on radiation in gravitational shockwave collisions}
\vskip 0.4in
{\large Isabelle Blackstad$^a$, Himanshu Raj$^b$,  Raju Venugopalan$^{b,c,d}$} 
\vskip .2in

$^a$ {\it Department of Physics and Astronomy, Vanderbilt University, Nashville, TN 37240,
USA}\\[0.5cm]

$^b$ {\it Center for Frontiers in Nuclear Science, Department of Physics and Astronomy,}\\
{\it Stony Brook University, Stony Brook, NY 11794, USA}\\[0.5cm]

$^c${\it  Department of Physics, Brookhaven National Laboratory,}
{\it  Upton, NY 11973, USA} \\[0.5cm]

$^d$ {\it Higgs Centre for Theoretical Physics, The University of Edinburgh,} 
{\it Edinburgh, EH9 3FD, UK}
 
\begin{abstract}
We compute high frequency gravitational wave radiation in shockwave collisions of a light mass compact object with energy $\mu_L$ off a heavy compact object (black hole) with energy $\mu_H \gg \mu_L$. A formal solution is obtained in terms of shockwave propagators that provide snapshots of $O(\mu_L\mu_H^n)$ tidal response functions sensitive to classical and potentially semi-classical dynamics of black hole formation. In a perturbative expansion, the first nontrivial contribution of $O(\mu_L\mu_H)$ is the emission of a graviton from the Lipatov vertex generated by the fusion of two reggeized gravitons. For graviton frequencies $\omega\rightarrow 0$, the  Lipatov amplitude reduces to the ultrarelativistic limit of the single-graviton exchange Weinberg amplitude. At $O(\mu_L\mu_H^2)$, the Lipatov amplitude includes a piece accounting for the rescattering of the graviton off a reggeized graviton and another, corresponding to graviton emission from a vertex fusing three reggeized gravitons. These nontrivial contributions again simplify to the Weinberg radiative amplitude for two-graviton exchange in the soft limit. We highlight throughout double copy structures to gluon radiation in dilute-dense collisions of Yang-Mills shockwaves. We relate our shockwave results to the  eikonal multiple scattering framework of Veneziano et al.  

\end{abstract}
\end{center}

\baselineskip=16pt
\setcounter{equation}{0}
\setcounter{footnote}{0}

\newpage
\tableofcontents

\section{Introduction}

Trans-Planckian $2\rightarrow 2+n$ scattering in Einstein gravity (GR) in high energy Regge asymptotics demonstrates a remarkable double copy to its counterpart in perturbative QCD~\cite{Lipatov:1982it,Lipatov:1982vv,Amati:1987uf,Amati:1990xe,Amati:1993tb}. The latter is described by the BFKL framework for $2\rightarrow 2+n$ scattering in multi-Regge kinematics~\cite{Kuraev:1976ge,Balitsky:1978ic,Lipatov:1991nf} whose building blocks are the Lipatov vertices governing the emission of gluons and reggeized gluon t-channel propagators. An identical framework applies in GR, where the gravitational Lipatov vertex satisfies the double copy relation 
\begin{equation}
\Gamma^{\mu\nu} = \frac{1}{2}C^\mu C^\nu - \frac{1}{2}N^\mu N^\nu\,. 
\label{eq:lipatov_dc}
\end{equation}
Here $C^\mu$ is the QCD Lipatov vertex governing $2\to 3$ gluon scattering in multi-Regge kinematics, and $N^\mu$ corresponds to the QED bremsstrahlung (soft photon) vertex up to an overall kinematic factor. The reggeized propagator in gravity similarly satisfies a double copy relation.  

A significant difference between perturbative QCD and GR however is that the evolution of the $2\rightarrow 2+n$ scattering amplitude with the center of mass energy $\sqrt{s}$ in the latter (described by the gravitational BFKL equation) is much weaker than in the QCD case. Specifically, the correction to the energy dependence of the Born amplitude ($\propto s^2$) due to BFKL evolution in GR is a $O(1/N)$ effect (with negative sign!), where $N$ denotes the very large occupancy of gravitons~\cite{Lipatov:1982vv,Bartels:2012ra}. A recent review of BFKL dynamics in QCD and in GR, which discusses in detail subtle features of the construction, is given in \cite{Raj:2025hse}. (See also \cite{Rothstein:2024nlq,Alessio:2025isu,Alessio:2026bdi} for related work.) 

QCD in the high energy Regge regime nevertheless has important lessons for GR that are especially relevant in the strong field regime. 
As summarized in \cite{Raj:2025hse}, BFKL evolution in rapidity drives QCD dynamics towards a high occupancy state of saturated gluons characterized by an emergent semi-hard scale $Q_S$, where $Q_S
\gg \Lambda_{\rm QCD}$. This is the physics of the Color Glass Condensate (CGC) EFT~\cite{Iancu:2003xm,Gelis:2010nm} whose ingredients are static (on the lightcone) classical color charge densities and dynamical gauge fields~\cite{McLerran:1993ni,McLerran:1993ka}. In the CGC EFT, high energy evolution of scattering amplitudes is described in terms of quark and gluon shockwave propagators~\cite{McLerran:1994vd,Ayala:1995kg,Balitsky:1995ub}, resulting in the BK/JIMWLK renormalization group equations~\cite{Balitsky:1995ub,Kovchegov:1999ua,Kovchegov:1999yj,Jalilian-Marian:1997ubg,Jalilian-Marian:1998tzv,Iancu:2000hn,Ferreiro:2001qy}. 
In this gluon saturation regime, the power counting for multiparticle production is modified qualitatively relative to perturbative QCD in Regge asymptotics, with the leading contribution in the CGC EFT arising from solutions of the Yang-Mills equations in the shockwave background~\cite{Kovner:1995ja,Kovchegov:1997ke,Krasnitz:1998ns,Krasnitz:1999wc,Krasnitz:2000gz,Lappi:2006fp,Gelis:2006cr,Gelis:2006dv,Berges:2020fwq}. As we will see, it is the CGC that has the remarkable double copy structure to the trans-Planckian strong field regime of gravitational radiation. Many of these conclusions are presaged in Lipatov's reggeon field theory for QCD and gravity~\cite{Lipatov:1982it,Lipatov:1982vv,Lipatov:1991nf}, though the approaches followed appear to be quite different.

Shockwave scattering in QCD \cite{Kovner:1995ja,Kovner:1995ts,Kovchegov:1997ke}, in general, can only be solved numerically because the Yang-Mills fields are nonperturbatively large, $A_\mu\sim 1/g$. However, one can identify parameters $\rho_L/\nabla_\perp^2, \rho_H/\nabla_\perp^2$ (where $\rho_L$, $\rho_H$ are the aforementioned color charge densities, respectively, of the colliding nuclei) that one can expand the YM equations in to obtain analytic solutions valid in specific kinematic contexts. In CGC jargon, these are the dilute-dilute  $\rho_A/\nabla_\perp^2, \rho_B/\nabla_\perp^2 \ll 1$~\cite{Kovner:1995ja,Kovner:1995ts,Kovchegov:1997ke,Gyulassy:1997vt} and dilute-dense   $\rho_L/\nabla_\perp^2\ll 1, \rho_H/\nabla_\perp^2 \sim 1$~\cite{Dumitru:2001ux,Blaizot:2004wu,Gelis:2005pt} limits.  In the dilute-dense approximation, the gluon radiation field was computed to be~\cite{Blaizot:2004wu,Gelis:2005pt}: 
\begin{align}
\label{eq:dilute-denseQCD}
a_i(k)  =  -\frac{2ig}{k^2+i\epsilon k^-}\int \frac{d^2\bsq_{2}}{(2\pi)^2} \(q_{2i}-k_i\frac{\bsq_2^2}{\bsk^2}\) \frac{\rho_L(\bsq_{2})}{\bsq_{2}^2}\bigg({\tilde U}(\bsk+\bsq_{2})-(2\pi)^2 \delta^2(\bsk+\bsq_{2})\bigg)\,,
\end{align}
where the term in the first parenthesis is the QCD Lipatov vertex $C^i$ in lightcone gauge. In the second parenthesis, $\tilde U$ is the Fourier transform of of the Wilson line 
\be
\label{QCD-shockwave}
    U(x^-, \bsx)  
    = \exp\(ig \int_{-\infty}^{x^-} dz^- \bar{A}_-(z^-, \bsx) \cdot T \)\,\,{\rm with}\,\,
\bar{A}_\m(x^-,\bsx) = -g \delta_{\mu-} \delta\left(x^{-}\right) \frac{\rho_{H}\left(\boldsymbol{x}\right)}{\nabla_\perp^2}~,
\ee
the background shockwave (``reggeon") gauge field. The Wilson line captures the coherent multiple scattering of the emitted gluon off the reggeized gluons in the dense source $\rho_H$; expanding this result to lowest order in $\rho_H$  recovers the dilute-dilute result. 
The fully nonperturbative dense-dense limit ($\rho_L/\nabla_\perp^2, \rho_H/\nabla_\perp^2 \sim 1$) corresponds to the scattering of two equally heavy nuclei with similar color charge densities; the numerical solutions~\cite{Krasnitz:1998ns,Krasnitz:1999wc,Krasnitz:2000gz,Lappi:2003bi,Berges:2013eia,Berges:2014yta} describe the evolution of QCD matter as it thermalizes to form a quark-gluon plasma~\cite{Berges:2020fwq}. 

\begin{figure}[ht]
    \centering
    \includegraphics[scale=1]{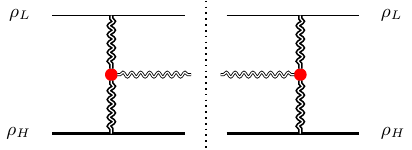}
    \caption{The ACV H-diagram, whose imaginary part gives the leading shockwave contribution to gravitational radiation. The dark wavy lines denote t-channel reggeized gravitons while the light wave lines represent the emitted graviton. The red blob is the gravitational Lipatov vertex in Eq.~\eqref{eq:lipatov_dc}. The transverse mass densities of the ultrarelativistic sources are labeled as $\rho_L$ and $\rho_H$.}
    \label{fig:PM-shockwave-G3}
\end{figure}

The quantitative connections between shockwave scattering in QCD and in GR were discussed by us at length previously\footnote{We further demonstrated in \cite{Raj:2023iqn} that this result can be obtained by applying a  classical color-kinematic duality~\cite{Goldberger:2016iau,Monteiro:2014cda} to the gluon shockwave result.} ~\cite{Raj:2023irr,Raj:2024xsi,Raj:2025hse,Stasto:2026xqw}. In particular, our understanding of the former allowed us to derive Eq.~\eqref{eq:lipatov_dc} for the first time in shockwave collisions. Our computation was performed in the dilute-dilute approximation ($\rho_L/\nabla_\perp^2, \rho_H/\nabla_\perp^2 \ll 1$) to Einstein's equations, with color charge densities replaced in this case by transverse mass densities. Specifically, in the shockwave limit, the energy of a light (heavy) colliding object  $\mu_{L(H)}= m_{L(H)}\gamma$ (where $m_{L(H)}$ is their mass and $\gamma$ the Lorentz factor in the center-of-frame) is kept fixed with $m_{L(H)}\rightarrow 0$ and $\gamma\rightarrow \infty$~\cite{Dray:1984ha,tHooft:1987vrq}. 
Fig.~\ref{fig:PM-shockwave-G3} depicts this dilute-dilute contribution to gravitational radiation, where the red blob denotes the gravitational Lipatov vertex in Eq.~\eqref{eq:lipatov_dc}.

As noted, trans-Planckian scattering\footnote{This is for the kinematical regime  
$b > R_S > \lambda_s$, where $b$ is the impact parameter, $R_S$ is the Schwarzschild radius, and $\lambda_s$ is the string length.} was studied previously in the scattering amplitude formalism by Lipatov, and by Amati, Ciafaloni and Veneziano  (ACV)~\cite{Amati:1987uf,Amati:1990xe,Amati:1993tb} within the framework of a 2+1-D reggeon EFT. Our leading order dilute-dilute shockwave computation is precisely the  imaginary part of the relevant ``H-diagram" computed by ACV in the Regge limit; this corresponds to $O(G^3)$ contributions to gravitational wave radiation in the post-Minkowski (PM) expansion \cite{Amati:1990xe, Damour:2020tta, DiVecchia:2021bdo, DiVecchia:2021ndb, Herrmann:2021tct, Herrmann:2021lqe,Bjerrum-Bohr:2021din}, where $G$ is Newton's constant.  As noted by ACV, the H-diagram provides the leading absorptive contribution to the eikonal gravitational S-matrix.

The dilute-dilute approximation in GR however does not fully capture the high frequency regime of  gravitational wave emission. Even with impact parameter $b\gg R_S$, an out-going high-frequency graviton produced close to one of the compact objects will experience a strong tidal response which, as discussed, can be interpreted as coherent rescattering in the QCD shockwave framework. An example of this phenomenon in GR would be gravitational radiation in extreme mass ratio inspirals (EMRIs) for the scenario where the radiation is emitted close to the heavier compact (black hole) object. It is interesting to consider how the tidal response is imprinted on this radiation in the shockwave limit and whether the corresponding response functions, in analogy to deeply inelastic scattering in the QCD case, can provide insight into black hole formation.

\begin{figure}[ht]
    \centering
    \includegraphics[scale=1]{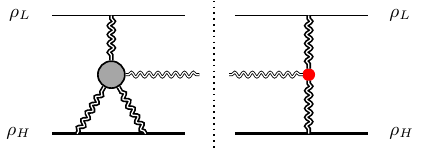}\hfill
    \includegraphics[scale=1]{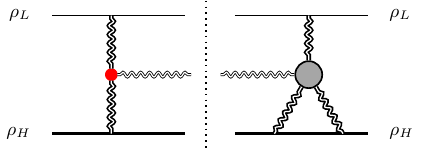}
    \caption{The $O(\mu_L^2\mu_H^3)$ contributions to the gravitational radiation spectrum.}
    \label{fig:PM-shockwave-G4}
\end{figure}
\begin{figure}[ht]
    \centering
    \includegraphics[scale=1]{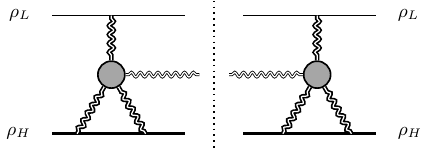}
    \caption{The $O(\mu_L^2\mu_H^4)$ contributions to the gravitational radiation spectrum.}
    \label{fig:PM-shockwave-G5}
\end{figure}

\begin{figure}[ht]
\centering
$$
\vcenter{\hbox{
    \includegraphics[width=0.2\textwidth]{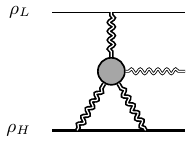}
}}
\;=\;
\vcenter{\hbox{
    \includegraphics[width=0.2\textwidth]{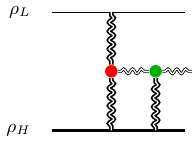}
}}
\;+\;
\vcenter{\hbox{
    \includegraphics[width=0.2\textwidth]{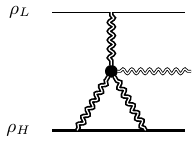}
}}
$$
\caption{Decomposition of the effective-vertex into the rescattering contribution containing the Lipatov vertex and the intrinsic three reggeon-graviton contribution.}
\label{fig:effective-vertex-decomposition}
\end{figure}

In this work, we will obtain a formal solution for the shockwave amplitude in  gravitational scattering to all-orders $O(\mu_L\mu_H^n)$ in the dilute-dense approximation. We are however at present only able to obtain explicit expressions for the $O(\mu_L \mu_H^2)$ contributions, which we obtain by employing a perturbative expansion of the all-order result. The contributions to the power spectrum are illustrated in Figs.~\ref{fig:PM-shockwave-G4} and \ref{fig:PM-shockwave-G5}. These are, respectively, $O(\mu_L^2\mu_H^3)$ and $O(\mu_L^2 \mu_H^4)$ in the power spectrum, and correspond likewise, to $O(G^4)$ and $O(G^5)$ in the PM expansion.  Both the Lipatov vertex and reggeized graviton propagators are key elements in this result. However unlike the QCD result in Eq.~\eqref{QCD-shockwave}, a novel structure arises at this order. As shown in Fig.~\ref{fig:effective-vertex-decomposition}, this structure has a piece that can be interpreted as the rescattering of a graviton emitted from the Lipatov vertex off a reggeon and an additional piece that corresponds to an intrinsic three reggeized graviton-graviton emission vertex. 

The rescattering contributions to gravitational wave radiation have also been discussed previously by Veneziano and collaborators~\cite{Ciafaloni:2015vsa,Ciafaloni:2015xsr,Gruzinov:2014moa,Ciafaloni:2018uwe}; we will consider in particular the detailed discussion by Ciafaloni, Coradeschi, Colferai and Veneziano (CCCV). A key result in understanding the relation of CCCV to our work is the equivalence between the leading eikonal contribution to the $2\rightarrow 2$ scattering amplitude~\cite{Kabat:1992tb,Amati:1992zb} and the semi-classical shockwave formalism~\cite{tHooft:1987vrq,Verlinde:1993mi}- see also an interesting discussion of this equivalence in the double copy context~\cite{Akhoury:2013yua}. As discussed by CCCV, multiple scattering contributions to gravitational wave radiation qualitatively alter the power spectrum. In the ACV formalism, these interpolate between the classical result of Weinberg~\cite{Weinberg:1965nx} for soft radiation and semi-hard Lipatov/Regge radiation in the presence of multiple scattering.  However in the shockwave formalism, as we showed explicitly in \cite{Raj:2025hse}, the  ultrarelativistic limit of the Weinberg current is already contained in the soft limit of the Lipatov H-diagram result at $O(\mu_L\mu_H)$. While multiple scattering contributions in the Lipatov regime cannot be exponentiated straightforwardly, this is possible in the soft limit, as shown by Weinberg. We will show explicitly that the Weinberg result for gravitational radiation at $O(\mu_L \mu_H^2)$ is also recovered smoothly at this order from the Lipatov result, providing an important test of our shockwave computation. The  discrepancy with CCCV may be a consequence of coherence effects in the emission that may be unimportant in the strict shockwave limit; this needs to be better understood. 

Our formulation of shockwave scattering in the language of the mass distributions $\rho_{L (H)}$ of the ultrarelativistic light (heavy) compact objects allows us to connect the tidal effects that contribute to the Lipatov radiation spectrum to the dynamical response of sources. In particular, in the dilute-dense EMRI scenario envisaged, the radiation spectrum will encode information on dynamical Love numbers~\cite{Rodriguez:2026iot,Binnington:2009bb,Damour:2009vw} that encapsulate the classical (and, in principle, quantum) spacetime response of the source which can be expressed in the shockwave formalism as the correlators $\langle \rho_H(x_1)\rho_H(x_2)\cdots\rho_H(x_n)\rangle$ of the mass/energy densities.\footnote{An important clarification is that in the shockwave limit what is being observed in these correlators is a snapshot of the tidal response at a given gravitational frequency. This is analogous to the deeply inelastic scattering case in QCD we mentioned previously. However, unlike QCD, the response functions in GR are sensitive to the effect of Shapiro time delay.}. As in worldline effective field theory~\cite{Goldberger:2005cd,Goldberger:2004jt,Cheung:2023lnj} the shockwave operators are universal gauge invariant objects, whose short distance coefficients should be interpreted as the corresponding Love numbers. They can in principle be extracted from measurements of Lipatov radiation, where the relation of the response to these many-body current-current correlation functions can be interpreted in the language of Raman scattering~\cite{Ivanov:2024sds,Ivanov:2026icp}. 

Our manifestly classical shockwave formalism is close in spirit to the self-force and worldline formalisms, which too can separately be matched to the scattering amplitude framework~\cite{Barack:2018yvs,Barack:2023oqp,Haddad:2025cmw}. It will also be interesting to match our all-order formal dilute-dense $O(\mu_L\mu_H^n)$ result to the path integral formulation in the BH perturbation theory worldline framework in \cite{Bohnenblust:2026ujk}. A useful exercise in particular will be to compute Lipatov radiation in the worldline framework. 

The paper is organized as follows. In Sections~\ref{sec:Definitions} and \ref{sec:Rad-dilute-dense-setup}, we will reintroduce the semi-classical shockwave framework  of \cite{Raj:2023irr} for solving the linearized Einstein equations in the Aichelburg-Sexl shockwave background, and write down Einstein's equations in the dilute-dense approximation corresponding to all orders in powers of  $\rho_H/\nabla_\perp^2$ but  fixed first order in $\rho_L/\nabla_\perp^2$, or equivalently, in the respective powers of $\mu_H$ and $\mu_L$. An important element in the subsequent solution is the description of the temporal evolution of the energy-momentum tensor of the heavy and light compact sources in Sec.~\ref{sec:EMTensor}. That of the heavy source is taken to be unchanged in the collision. In contrast, the temporal evolution of the energy-momentum tensor of the light source is  modeled as the (null) geodesic evolution of a collection of point-particle sources. With this input,  one can then write down the equations for the gravitational radiation spectrum in terms of the source densities to all orders in the dilute-dense approximation. 

In  Sec.~\ref{sec:exact-solution}, we obtain the formal all-order solution to gravitational radiation in the dilute-dense approximation. This is done employing the Green function method, where the dressed shockwave propagator can be expressed in terms of the free propagator in lightcone gauge, and an effective vertex, which was computed previously in \cite{Raj:2024xsi}. As discussed there, this effective vertex is proportional to the gravitation Wilson line containing all orders in $\mu_H$. As also observed in \cite{Raj:2024xsi}, the shockwave propagator is the GR double copy of shockwave propagators in QCD. A simple closed form solution is however difficult to achieve due to the the complexity of the source terms in the all-order expansion. 

In Sec.~\ref{sec:perturbative}, we show that equations for the gravitational field can be expressed as a hierarchy of equations, and solved perturbatively for each successive order in $O(\mu_L \mu_H^n)$. At the lowest order ($\mu_H^0$, we see in Sec.~\ref{sec:zeroth-order} that the solution is simply the Aichelberg-Sexl field. In Sec.~\ref{sec:Lipatov-vertex-calculation}, we recover our previous result for the gravitational Lipatov vertex $\Gamma_{ij}$. A novel feature of our derivation here is that a contact term introduced in \cite{Raj:2023irr} in the shockwave computation to reproduce the $N^\mu N^\nu$ term in Eq.~\eqref{eq:lipatov_dc} (required by unitarity for the 5-point scattering amplitude) is understood to not be {\it ad hoc} but instead  required for the correct smearing of the relevant shockwave operators. 

We next review in Sec.~\ref{subsec:angular} the angular structure of Lipatov radiation previously studied in \cite{Ciafaloni:2015vsa,Ciafaloni:2015xsr,Ciafaloni:2018uwe}. At $O(\mu_L\mu_H)$ within the shockwave framework, the contributions of the Weinberg and Lipatov regimes match smoothly in the frequency spectrum. The former correspond to radiative contributions to the 5-point eikonal scattering amplitude from the external legs and are 
$O(\omega^{-1})$ in the graviton frequency $\omega$. The latter includes, in addition to the Weinberg terms, the self-coupling of the exchanged graviton, as well as the emission of a soft graviton from the eikonal three-graviton vertex; these contributions are of order $O(\omega^0)$ and therefore subleading when $\omega\rightarrow 0$.

We then lay out in Section~\ref{sec:NLO} a systematic computation of the gravitational radiational field to $O(\mu_L\mu_H^2)$. The computation  is detailed in Secs.~\ref{sec:source-terms}-\ref{sec:complete-rescattering}.
There are two distinct contributions at this order: One can be interpreted as a rescattering contribution to the emission from the Lipatov vertex which the other is graviton emission from the fusion of three reggeized gravitons. Our final result for the radiative gravitational field at NLO is given in Eq.~\eqref{final-result}. The soft limit of the corresponding Lipatov amplitude is computed in Sec.~\ref{sec:Lipatov-soft} and is shown to match the high energy limit of the Weinberg amplitude at this NLO order as well. As noted, this is a nontrivial check of our result.  

Our result for the Lipatov amplitude at leading order suggests that the gravitational wave spectrum behaves as $1/\omega^2$ for large $\omega$. However the $O(\mu_L\mu_H^2)$ calculation reveals that the leading contribution to the amplitude at this order goes as $O(\omega)$ at large $\omega$, thereby giving a super-leading contribution to the spectrum. This result suggests that the all-order dilute-dense resummation is essential before we can address questions regarding the UV sensitivity of the spectrum raised in \cite{Gruzinov:2014moa}. We will see further that this sensitivity also depends on the correlators of $\rho_H$, potentially encoding dynamics at the scale of the Schwarzchild radius. In Sec.~\ref{sec:comments}, we discuss the breakdown in the double copy that is seen at $O(\mu_L\mu_H^2)$. Nevertheless, some features of the all-order exponentiation seen in the QCD case  persist, leaving open the possibility that an exponentiation of the leading order result may survive in some form. 

We conclude in Section~\ref{sec:Conclusions} with a summary of our results and a discussion of further work. The paper includes two  Appendices. In Appendix~\ref{app:CCV}, we discuss in some detail the results of the CCCV eikonal multiple scattering formalism of the  matching of the Regge and Weinberg regimes . We note that this matching has also been discussed recently in an EFT framework in \cite{Saavedra:2026ttk}. As we will discuss, it will be important to understand better the connections of these works to ours. In Appendix~\ref{app:eikonal-soft-NLO}, we provide an alternative derivation within the eikonal scattering formalism, of the soft (Weinberg) limit of Lipatov radiation in  dilute-dense scattering to $O(\mu_L\mu_H^2)$. 

\newpage
\section{Shockwave collisions in the dilute-dense framework}
\label{sec:setup}
In this section, we will review the gravitational shockwave formalism in a coordinate frame (gauge choice) where the metric takes a simple form and derive the equation for the gravitational radiation produced in the collision of two such shockwaves. 
\subsection{The shockwave background: Definitions}
\label{sec:Definitions}
Consider Einstein's equations
\begin{equation}\label{fieldEq1}
R_{\mu \nu}-\frac{1}{2} g_{\mu \nu} R=\kappa^2 T_{\mu \nu}~,\qquad \kappa^2 = 8\pi G~,
\end{equation}
in the presence of the EM tensor at initial times ($t=-\infty$) that generates the shockwave background:
\be
\label{initial-T-1}
T_{\H,--}=\mu_H\delta(x^-)\rho_H(\bsx)~,
\ee
where the subscript $H$ labels the shockwave as ``heavy" or ``dense". Here $\rho_H(\bsx)$ is the density in the transverse space of the shockwave, with $\mu_H \rho_H$ corresponding to the energy density. Further, the $\delta$-function in $x^-$ (with lightcone coordinates defined as 
$x^\pm = (t\pm z)/\sqrt{2}$), corresponds to the singular limit 
where the Lorentz factor $\gamma\rightarrow \infty$, mass $m_H\rightarrow 0$, and energy $\mu_H = m_H \gamma$ is kept fixed. 

For this EM tensor, the Einstein equations can be solved with the  ansatz 
\be\label{bgnd1}
\bar{g}_{\m\n}(x^-,\vec{x})=\left(
\begin{array}{cccc}
 0 & 1 & 0 & 0 \\
 1 &  \bar{g}_{--}(x^-,\bsx) & 0 & 0 \\
 0 & 0 & -1 & 0 \\
 0 & 0 & 0 & -1 \\
\end{array}
\right)\,,
\ee
with the shockwave metric $\bar{g}_{--}(x^-,\bsx)$ determined to be\footnote{Note the similarity in the structure of this metric to the shockwave gauge field in Eq.~\eqref{QCD-shockwave}.} 
\begin{align}
\label{bgnd1-sol}
\bar{g}_{--}(x^-,\bsx)  &=2\kappa^2 \mu_H\delta(x^-)\frac{\rho_H(\bsx)}{\square_\perp}  ~.
\end{align}
The solution can be expressed in terms of the line element as
\be
\label{bgnd-2}
ds^2 = 2dx^+ dx^- +  \bar{g}_{--}(x^-,\bsx)  \(dx^-\)^2 -\delta_{ij} dx^idx^j~.
\ee
Eqs.~\eqref{bgnd1} and \eqref{bgnd1-sol} constitute an exact solution (to all orders in $\rho_H$) of the fully nonlinear Einstein equations. 

There are of course other gauge choices (coordinate frames) where $\rho_H$ appears nonlinearly in the exact solution. One such gauge choice is obtained by performing the coordinate transformation
\begin{align}
\label{xy-transform}
\begin{split}
x^- &= y^-~,\\
x^+ &= y^+ -\Theta(y^-) \frac{1}{\square_\perp}\rho_H(\bsy) +\frac12  y^-\Theta(y^-) \(\frac{\p_i}{\square_\perp} \rho_H(\bsy)\)^2~,\\
x^i &= y^i - y^- \Theta(y^-)\frac{\p_i}{\square_\perp} \rho_H(\bsy) ~.
\end{split}
\end{align}
Recalling that under a general coordinate transformation $x^\m\to y^\m$, the metric transforms as 
\be
g_{\m\n}(y) = \frac{\p x^\rho }{\p y^\m } \frac{\p x^\sigma }{\p y^\n } g_{\rho\sigma}(x)~,
\ee
with the explicit result in ``y-coordinate" frame: 
\begin{align}\label{bgndC0}
\begin{split}
&\bar{g}_{++}=\bar{g}_{--}=\bar{g}_{+i}=\bar{g}_{-i}=0~,~~~~\bar{g}_{+-}=1~,\\[5pt]
&\bar{g}_{ij}= -\delta_{ij}+y^-\Theta(y^-)\[2\p_i\p_j \frac{1}{\square_\perp}\rho_H(\bsy)-y^- \p_i\p_k \(\frac{1}{\square_\perp}\rho_H(\bsy)\) \p_j\p_k \(\frac{1}{\square_\perp}\rho_H(\bsy)\)  \]~.
\end{split}
\end{align}
In this form, the metric has no discontinuity but is quadratic in the source density $\rho_H$. It too is an {\it exact} solution to Einstein's equations for the stress tensor given in Eq.~\eqref{fieldEq1}. Indeed, one can verify via an explicit calculation in both frames, that all the components of the Riemann tensor vanish for both $x^->0$ and $x^-<0$. While the above form is significantly more complicated when compared to Eq.~\eqref{bgnd1}, it makes manifest the fact that the vacua on the two sides of the shockwave ($x^->0$ and $x^-<0$) are distinct Minkowski vacua. Interestingly, this feature of the gravitational shockwave is identical to the CGC shockwave construction in QCD~\cite{McLerran:1993ni,McLerran:1993ka}, where the field strength tensor vanishes outside the shockwave. Further, in Lorenz gauge, the shockwave gauge field has an identical structure to Eq.~\eqref{bgnd1-sol}, while it has a nonlinear dependence on color charge densities in lightcone gauge~\cite{Ayala:1995kg,Balitsky:1995ub,Iancu:2003xm}. 

In order to perform any kind of shockwave computation (be it of gauge invariant or gauge variant quantity), we find it is most convenient\footnote{This again mirrors our experience in QCD, where the form of shockwave propagators is simpler in the singular Lorenz gauge~\cite{McLerran:1994vd,Balitsky:1995ub}.
} to work with the form of the metric in Eq.~\eqref{bgnd-2}. If needed one can transform the result to the frame Eq.~\eqref{bgndC0} via the transformation in Eq.~\eqref{xy-transform}. 

\subsection{Gravitational radiation in the dilute-dense approximation}
\label{sec:Rad-dilute-dense-setup}
Having computed the background shockwave metric of the heavy source, we will now consider small fluctuations around this background. To do so, we expand the metric as
\be
\label{grav-fluctuation}
g_{\m\n}= \bar{g}_{\m\n}+h_{\m\n}~,
\ee
and substitute it into the Einstein equations in Eq.~\eqref{fieldEq1}.
Since, as emphasized earlier, $\bar{g}_{\m\n}$ in Eq.~\eqref{bgnd1-sol} is to all orders in the dense source $\rho_H$, so too therefore are the small fluctuations $h_{\m\n}$. We distinguish two cases: 
\begin{enumerate}
    \item $h_{\m\n}$ are source-less excitations, an example being a passing gravitational wave.
    \item $h_{\m\n}$ is produced by another source. This would be the case when there are other sources contributing to the energy-momentum tensor such as a compact ``light" component $T_{L,\m\n}$ component in addition to Eq.~\eqref{initial-T-1}. This would for instance correspond to the EMRI scenario noted in the introduction. 
\end{enumerate}
We are here interested in the second case. To be specific, we take $T_{\L, \m\n}$ at initial times ($t=-\infty$) to be an ultrarelativistic source generating a shockwave traveling in the opposite lightcone direction to $T_{\H, \m\n}$. It is given by
\begin{align}
    \label{initial-T-2}
    T_{\L, \m\n} = \delta_{\m,+}\delta_{\n,+} \mu_L\delta(x^+)\rho_L(\bsx)~,
\end{align}
where $\rho_L(\bsx)$ is a localized transverse density profile that we assume to be separated from the center of $\rho_H(\bsx)$ by impact parameter\footnote{In the point-particle approximation, $\rho_L(\bsx)=\delta^{(2)}(\bsx-\bsb)$.} $\bsb$. The total energy-momentum tensor of the matter system at initial times is given by 
\begin{align}
\label{initial-EM-tensor}
    T_{{\rm init.},\mu\nu} = \delta_{\m,-}\delta_{\n,-}\mu_H\delta(x^-)\rho_H(\bsx)+\delta_{\m,+}\delta_{\n,+}\mu_L\delta(x^+)\rho_L(\bsx)~\,,
\end{align}
where the first term corresponds to $T_{\H, \m\n}$.
The goal here is to determine  the final (post-shockwave collision) energy-momentum-tensor $T_{\rm{final}, \m\n}$ of the matter system and the produced gravitational field $h_{\m\n}(x)$ to all orders in $\rho_H$ (or equivalently, $\mu_H$) and to first order in $\rho_L$ ($\mu_L$). The initial condition of this collision problem is completely specified by the energy-momentum tensor above. 

Plugging Eq.~\eqref{grav-fluctuation} in Einstein's equations, and fixing lightcone gauge $h_{\m+} = 0~$, the gravitational radiation field equation in the dilute-dense approximation takes the form\footnote{See \cite{Raj:2023irr} for the details of this derivation.}:

\begin{align}
\label{dilute-dense-radiation-eom}
 \bar{g}_{--}\p_+^2 \tilde h_{ij}-2\p_+\p_-\tilde h_{ij} + \square_\perp \tilde h_{ij}=& \kappa^2\bigg[\(2\p_i\p_j-\square_\perp \delta_{ij}\)\frac{1}{\p_+^2}T_{++}+2T_{ij}-\delta_{ij}T\no\\[5pt]
&-\frac{2}{\p_+} \(\p_iT_{+j}+\p_jT_{+i}-\delta_{ij}\p_k T_{+k}\)\bigg] ~.
\end{align}
Here $\square_\perp$ is the transverse derivative in flat space, and the energy-momentum tensor components are those of the entire dilute-dense matter system for all times (both pre- and post-collision). Further, $T=\delta_{ij}T_{ij}$. In order to solve this equation, we first need to know the energy-momentum tensor. Following \cite{Raj:2023irr}, we will now determine these energy-momentum tensor sources in the point-particle approximation, and generalize it subsequently. 

\subsection{EM tensor evolution in the point-particle approximation}
\label{sec:EMTensor}

The EM tensor of the light shockwave at initial times in the point-particle approximation is
\begin{align}
\label{initial-EM-tensor-1}
    T_{\rm {init.L}, \mu\nu} (x)= \delta_{\m,+}\delta_{\n,+}\mu_L\delta(x^+)\delta^{(2)}(\bsx-\bsb)~.
\end{align}
Treating the light shockwave as a single particle is a mathematical abstraction since in quantum field theory this particle is surrounded by a cloud of gravitons that, together with the point-particle, forms the smeared transverse density $\rho_L(\bsx)$. We will restore this smeared form towards the end of the calculation of the components of the EM tensor that we first perform using the point-particle approximation.

At negative times, the ``light" particle is at impact parameter $\bsb$ in the transverse space relative to the heavy shockwave; note that  we still take the latter to be a dense collection of particles localized at the origin in transverse space.
Therefore, the initial ($t<0$) energy-momentum tensor is
\begin{align}
    T_{{\rm init.},\m\n}(x)  = T_{{\rm init.}, \L,\m\n}  (x)+T_{{\rm init.}, \H,\m\n} (x)
\end{align}
where
\begin{align}
    &T_{\rm{init. L},\m\n}(x) = \delta_{\m,+}\delta_{\n,+}\mu_L \delta(x^+)\delta^{(2)}(\bsx-\bsb)~,\no\\[5pt]
    &T_{\rm{init. H},\m\n}(x) = \delta_{\m,-}\delta_{\n,-}\mu_H \delta(x^-)\rho_H(\bsx)~.
\end{align}
We wish to determine how the above initial energy-momentum tensor is modified in the post collision region $t>0$.

In the dilute-dense approximation ($\mu_H\gg \mu_L$), we do not expect the EM tensor of the heavy compact source to be modified by the light particle\footnote{While we will ignore the overall center-of-mass motion of this heavy source induced by the recoil in the scattering off the light particle, we will later discuss the correlations in its matter distribution induced by gravitational radiation occuring at smaller impact parameters.}. That is, we have
\begin{align}
\label{EM-final-H}
    T_{\rm{final\, H},\m\n}(x)= T_{\rm{init. H},\m\n}(x)~.
\end{align}

Our task is therefore to determine the evolution of $T_{final, \L,\m\n}(x)$ alone. To compute this, we will employ the general formula for the point-particle energy-momentum tensor~\cite{Weinberg:1972kfs},
\begin{align}
\label{PPEMT}
T^{\m\n}(x) = \frac{\mu_L}{\sqrt{-g(x)}} \int_{-\infty}^\infty d\lambda \frac{dX^\m(\lambda)}{d\lambda}\frac{dX^\n(\lambda)}{d\lambda} \delta^{(4)}(x-X(\lambda))~,
\end{align}
where $\lambda$ is the worldline parameter of the light particle and $g(x)$ is the determinant of the background metric Eq.~\eqref{bgnd1} created by the heavy shockwave, which is simply equal to $-1$. 

\subsubsection{Solution of the null geodesic equation}
Recall the geodesic equation, along with the null constraint,
\be
\label{geodesicEq}
\frac{d^2X^\mu(\lambda)}{d\lambda^2}+\Gamma^\mu_{\nu\rho} \frac{dX^\nu(\lambda)}{d\lambda}\frac{dX^\rho (\lambda)}{d\lambda}=0~,\qquad g_{\nu\rho}\frac{dX^\nu(\lambda)}{d\lambda}\frac{dX^\rho (\lambda)}{d\lambda}=0~.
\ee

As mentioned above, in order to calculate the energy-momentum tensor in Eq.~\eqref{PPEMT} we first need to need to solve the geodesic equation to  determine the  worldline $X^\m(\lambda)$ in the background of the dense shockwave given in Eq.~\eqref{bgnd-2}. The Christoffel symbols
\be
\Gamma^\mu_{\nu\rho}(x) = \frac12 g^{\m\lambda}\(\p_\nu g_{\lambda\rho}+\p_\rho g_{\nu \lambda}-\p_\lambda g_{\nu\rho}\)~,
\ee
for the metric in Eq.~\eqref{bgnd-2} have only the following nonvanishing components:
\be
\Gamma^{+}_{--}=\frac12 \partial_-\bar{g}_{--}~,\qquad \Gamma^{+}_{-i}=\Gamma^{i}_{--}=\frac12 \partial_i\bar{g}_{--}~.
\ee
The geodesic equations for this case therefore simplify to 
\begin{align}
&\ddot{X}^- = 0~,\qquad \ddot{X}^i + \Gamma^i_{--}\(\dot{X}^-\)^2 = 0~, \no\\
&\ddot{X}^+ + \Gamma^+_{--}\(\dot{X}^-\)^2 + 2\Gamma^+_{-i}\dot{X}^-\dot{X}^i = 0 \,,
\end{align}
where the dot superscripts denote differentiation w.r.t. to the worldline affine parameter $\lambda$. Without loss of generality,  
$X^-  = \lambda$, giving $\dot{X}^-=1$. Then 
\begin{align}
\begin{split}
    X^i(\lambda) &= c_2^i + c_1^i\lambda -\kappa^2 \mu_H \lambda \Theta(\lambda) \p_i\frac{\rho_H(X^i(0))}{\square_\perp}~,
\end{split}
\end{align}
where $c_2^i,c_1^i$ are constants of integration. The boundary condition $X^i(\lambda=-\infty) = b^i$ sets $c_1^i=0$ and $c_2^i=b^i$, with $X^i(0) = b^i$ in the argument of $\rho_H$. Hence, 
\begin{align}
    X^i(\lambda) = b^i -\kappa^2 \mu_H \lambda \Theta(\lambda) \p_i\frac{\rho_H(\bsb)}{\square_\perp}\,.
\end{align}
Likewise, integrating the null constraint in Eq.~\eqref{geodesicEq} with the boundary condition $X^+(\lambda=-\infty) = 0$, we get 
\be
\label{solXp-null-condition}
X^+(\lambda) = -  \kappa^2\mu_H \Theta(\lambda) \frac{\rho_H(\bsb)}{\square_\perp} +\frac{\kappa^4 \mu_H^2}{2}\lambda \Theta(\lambda) \(\frac{\p_i\rho_H(\bsb)}{\square_\perp}\)^2~.
\ee

In summary, the solutions to the geodesic equation are\footnote{Note that these null geodesic solutions are not the only ones in the shockwave spacetime of Eq.~\eqref{bgnd-2}. There is another class of null geodesics, which satisfy $\dot{X}^-=0$, $\dot{X}^i=d^i$, a constant everywhere, and $X^+ = \lambda$. This solution satisfies the geodesic equation and represent geodesics along the perpendicular lightcone.}
\begin{align}
\begin{split}
    X^-(\lambda) &= \lambda~,\qquad X^i(\lambda) = b^i -\kappa^2 \mu_H \lambda \Theta(\lambda) \p_i\frac{\rho_H(\bsb)}{\square_\perp}~,\\[10pt]
    X^+(\lambda) &= -  \kappa^2\mu_H \Theta(\lambda) \frac{\rho_H(\bsb)}{\square_\perp} +\frac{\kappa^4 \mu_H^2}{2}\lambda \Theta(\lambda) \(\frac{\p_i\rho_H(\bsb)}{\square_\perp}\)^2~.
\end{split}
\end{align}
Since we will also need their first derivatives, let us collect these as well:
\begin{align}
\begin{split}
    \dot X^-(\lambda) &= 1~,\qquad \dot X^i(\lambda) = -\kappa^2 \mu_H  \Theta(\lambda) \p_i\frac{\rho_H(\bsb)}{\square_\perp}~,\\[10pt]
    \dot X^+(\lambda) &= -  \kappa^2\mu_H \delta(\lambda) \frac{\rho_H(\bsb)}{\square_\perp} +\frac{\kappa^4 \mu_H^2}{2} \Theta(\lambda) \(\frac{\p_i\rho_H(\bsb)}{\square_\perp}\)^2~.
    \label{geodesic-sol}
\end{split}
\end{align}

\subsubsection{EM tensor in the dilute-dense approximation}
\label{sec:EMtensor-dilute-dense}
Now that we have computed the worldline trajectory of the light particle, we can substitute Eq.~\eqref{geodesic-sol} into Eq.~\eqref{PPEMT}; we obtain,
\begin{align}
    T^{\m\n}_{\L}(x^+, x^-, \bsx) &= \mu_L \dot{X}^\mu(x^-)\dot{X}^\nu(x^-) \delta(x^+-X^+(x^-))\delta^{(2)}\(\bsx-\bsX(x^-)\)\,,
\end{align}
which is the exact result for the time-evolved stress tensor of a light particle $L$ before and after its collision with the shockwave of the heavy compact source. For our subsequent discussion, it is worth our while to explicitly write out the various components of this expression:
\begin{subequations}
\begin{align}
    T_{\L}^{--} &=\mu_L \Delta~,\\[10pt]
    T_{\L}^{++} &=\mu_L \[-  \kappa^2\mu_H \delta(x^-) \frac{\rho_H(\bsb)}{\square_\perp} +\frac{\kappa^4 \mu_H^2}{2} \Theta(x^-) \(\frac{\p_i\rho_H(\bsb)}{\square_\perp}\)^2\]^2\Delta~,\\[10pt]
    T_{\L}^{+-} &=\mu_L\[-  \kappa^2\mu_H \delta(x^-) \frac{\rho_H(\bsb)}{\square_\perp} +\frac{\kappa^4 \mu_H^2}{2} \Theta(x^-) \(\frac{\p_i\rho_H(\bsb)}{\square_\perp}\)^2\] \Delta~,\\[10pt]
    T_{\L}^{-i} &=\mu_L \(-  \kappa^2\mu_H \Theta(x^-) \frac{\p_i\rho_H(\bsb)}{\square_\perp}\)\Delta~,\\[10pt]
    T_{\L}^{+i} &=\mu_L \[-  \kappa^2\mu_H \delta(x^-) \frac{\rho_H(\bsb)}{\square_\perp} +\frac{\kappa^4 \mu_H^2}{2} \Theta(x^-) \(\frac{\p_i\rho_H(\bsb)}{\square_\perp}\)^2\]\(-  \kappa^2\mu_H \Theta(x^-) \frac{\p_i\rho_H(\bsb)}{\square_\perp}\)\Delta~,\\[10pt]
    T_{\L}^{ij} &=\mu_L\kappa^4\mu_H^2 \Theta(x^-) \frac{\p_i\rho_H(\bsb)}{\square_\perp}\frac{\p_j\rho_H(\bsb)}{\square_\perp} \Delta~.
\end{align}
\end{subequations}
In these expressions, $\Delta$ is the product of delta functions,
\be
\label{Delta-def}
\Delta \equiv \delta\(\tilde{x}^+\)\delta^{(2)}\(\tilde{x}^i\)~,
\ee
where 
\begin{align}
\label{x-tilde-shorthand}
    \tilde{x}^+(x^+, x^-, \bsb) &\equiv x^+ + \kappa^2\mu_H \Theta(x^-) \frac{\rho_H(\bsb)}{\square_\perp} -\frac{\kappa^4 \mu_H^2}{2} x^-\Theta(x^-) \(\frac{\p_i\rho_H(\bsb)}{\square_\perp}\)^2~,\no\\
    \tilde{x}^i(x^i, x^-, \bsb) &\equiv x^i -b^i + \kappa^2\mu_H x^-\Theta(x^-) \frac{\p_i\rho_H(\bsb)}{\square_\perp}~.
\end{align}
We see explicitly therefore that the components of the energy-momentum tensor are sensitive 
to contributions to all orders in powers of $\rho_H$. With these expressions, we have now determined\footnote{As a final step, we will need to lower the tensor indices of all of these components of the EM tensor via the relation: $T_{\m\n} = g_{\m\alpha}g_{\n\beta} T^{\alpha\beta}$ and add the contribution $T_{\rm{final\, H},\m\n}$ in given in Eq.~\eqref{EM-final-H} for the EM tensor of the heavy quark source.} the r.h.s of the equation for the radiation field in Eq.~\eqref{dilute-dense-radiation-eom}.

\subsection{Solution for the radiation field in the dilute-dense approximation}
\label{sec:exact-solution}

To recapitulate the results from the previous sections, the differential equation governing the radiation field in the dilute-dense approximation can be written compactly as 
\begin{align}
\label{dilute-dense-radiation-eom-1}
 \bar{g}_{--}\p_+^2 \tilde h_{ij}-2\p_+\p_-\tilde h_{ij} + \square_\perp \tilde h_{ij}=& \kappa^2 S_{ij} ~,
\end{align}
where the matter source term is 
\begin{align}
\label{eq:Sij}
    S_{ij} & = \(2\p_i\p_j-\square_\perp \delta_{ij}\)\frac{1}{\p_+^2}T_{++}+2T_{ij}-\delta_{ij}T-\frac{2}{\p_+} \(\p_iT_{+j}+\p_jT_{+i}-\delta_{ij}\p_k T_{+k}\)\,.
\end{align}
The explicit expressions for the components of the energy-momentum tensor here are
\begin{align}
    T_{++}(x,b) &=\mu_L \Delta\,,\\[5pt]
    T_{+i}(x,b) &=\mu_L \(\kappa^2\mu_H \Theta(x^-) \frac{\p_i\rho_H(\bsb)}{\square_\perp}\)\Delta\,,\\[5pt]
    T_{ij}(x,b) &=\mu_L\kappa^4\mu_H^2 \Theta(x^-) \frac{\p_i\rho_H(\bsb)}{\square_\perp}\frac{\p_j\rho_H(\bsb)}{\square_\perp} \Delta~,
\end{align}
where $\Delta$ is defined in Eqs.~\eqref{Delta-def} and \eqref{x-tilde-shorthand}.

We will now derive the exact solution of the initial value problem posed by Eq.~\eqref{dilute-dense-radiation-eom-1}. The relevant retarded Green function $G_{i j k l}(x-y)$ in the shockwave background was obtained in \cite{Raj:2024xsi}. It satisfies the equation 
\be
2 \partial_{+} \partial_{-} G_{i j k l}(x, y)-\square_{\perp} G_{i j k l}(x, y)-g_{--} \partial_{+}^2 G_{i j k l}(x, y)=\frac{1}{2}\left(\delta_{i k} \delta_{j l}+\delta_{i l} \delta_{j k}-\delta_{i j} \delta_{k l}\right) \delta^{(4)}(x-y) .
\ee
The solution of this equation in momentum space is
\be
\label{green-function}
\tilde{G}_{i j k l}\left(p, p^{\prime}\right)=\tilde{G}_{i j k l}^0(p)(2 \pi)^4 \delta^{(4)}\left(p-p^{\prime}\right)+\tilde{G}_{i j p q}^0(p) \mathcal{T}^{p q r s}\left(p, p^{\prime}\right) \tilde{G}_{r s k l}^0\left(p^{\prime}\right)~,
\ee
where $\tilde{G}_{i j p q}^0(p)$ is the free graviton propagator 
\be
G_{i j k l}^0(x, y)=\frac{1}{2}\left(\delta_{i k} \delta_{j l}+\delta_{i l} \delta_{j k}-\delta_{i j} \delta_{k l}\right) G_R^0(x, y)~,
\ee
with
\be
G_R^0(x, y)=-\int \frac{d^4 k}{(2 \pi)^4} \frac{e^{-i k \cdot(x-y)}}{k^2+i \epsilon k^{-}}=\frac{1}{2 \pi} \Theta\left(x^{-}-y^{-}\right) \Theta\left(x^{+}-y^{+}\right) \delta\left((x-y)^2\right)~.
\ee
The ``effective vertex" tensor $\mathcal{T}_{\mu \nu \rho \sigma}\left(p, p^{\prime}\right)$ in Eq.~\eqref{green-function} is given by
\begin{eqnarray}
\label{T-tensor}
\mathcal{T}_{\mu \nu \rho \sigma}\left(p, p^{\prime}\right)&=&-\frac{1}{2}\left(\Lambda_{\mu \rho} \Lambda_{\nu \sigma}+\Lambda_{\mu \sigma} \Lambda_{\nu \rho}-\Lambda_{\mu \nu} \Lambda_{\rho \sigma}\right) 4 \pi i\left(p^{\prime}\right)^{-} \delta\left(p^{-}-\left(p^{\prime}\right)^{-}\right) \nonumber \\
&\times&\int d^2 \boldsymbol{z}\, e^{i\left(\boldsymbol{p}-\boldsymbol{p}^{\prime}\right) \cdot \boldsymbol{z}}\left(e^{i f_1(\boldsymbol{z}) p^{-}}-1\right)\,.
\end{eqnarray}
Here $f_1(\boldsymbol{x})=\kappa\mu_H \rho_H(\bsx)/\square_\perp$ and $\Lambda_{\mu \nu}=\eta_{\mu \nu}-\frac{n_\mu p_\nu+n_\nu p_\mu}{n \cdot p}$. 

These expressions for the retarded Green's functions are have an identical structure to fermion and gluon retarded shockwave Green functions in QCD~\cite{McLerran:1998nk,Balitsky:2001mr}. See for instance \cite{Roy:2019hwr,Caucal:2021ent} for an explicit demonstration of their use in deeply inelastic scattering computations in QCD. In particular, the expression on the second line is nothing but the Fourier transform of the gravitational Wilson line; the latter is a double copy of the expression one obtains in QCD~\cite{Melville:2013qca}. 

The solution of the initial value problem in Eq.~\eqref{dilute-dense-radiation-eom-1} can be formally written down in terms of the retarded Green function, as outlined in \cite{Raj:2024xsi}. This derivation too is exactly analogous to the QCD  discussion in \cite{Blaizot:2004wu}.  Given the initial value of the field $\bar{h}_{ij}$ at $x^-=x_0$, its value at $x^-> x_0$ is 
\begin{align}\label{tep0}
\begin{split}
\tilde{h}_{ij}(x)\bigg|_{x^->x_0} =-& \kappa^2\int_{y^->x_0} d^4y ~G_{ijkl}(x-y) S_{kl}(y)+\int_{y^-= x_0} dy^+ d^2 \boldsymbol{y}_{\perp}  G_{ijkl}(x-y)2\p_{y+}\tilde{h}_{kl}(y)~.
\end{split}
\end{align}
Since $G_{ijkl}$ in Eq.~\eqref{green-function} is known explicitly for the dilute-dense case, 
this result is an all-order (in $\mu_H$) formal solution of Eq.~\eqref{dilute-dense-radiation-eom-1}. Unlike the QCD dilute-dense result in Eq.~\eqref{QCD-shockwave} for the gluon radiation field, we do not have a similarly compact all-order solution.  
In what follows, we will analyze the formal result to $O(\mu_H^2)$ in a perturbative expansion.

\subsection{Perturbative expansion}
\label{sec:perturbative}
We expand both $\tilde{h}_{ij}$ and $T_{\mu\nu}$ in powers of $\mu_L$ and $\mu_H$:
\begin{equation}
\tilde{h}_{ij} = \sum_{a=0,1;\,b=0}^{\infty}\mu_L^a\mu_H^b\,\tilde{h}^{(a,b)}_{ij}\,, \qquad T_{\mu\nu} = \sum_{a=0,1;\,b=0}^{\infty}\mu_L^a\mu_H^b\,T^{(a,b)}_{\mu\nu}\,.
\label{eq:pert-exp}
\end{equation}
Since the light source contributes only at first order in $\rho_L$, the gravitational wave sector of interest corresponds to the $a=1$ term. As we discussed previously, keeping only the linear terms in $\rho_L$ is appropriate for physical situations such as EMRIs or for ultrarelativistic collisions where is gravitational wave is produced far from the light compact object but close to the heavy one. 

Substituting Eq.~\eqref{tep0} into Eq.~\eqref{dilute-dense-radiation-eom-1}, and matching powers of $\mu_L\mu_H^b$ for $b=0,1,2$, one obtains the following hierarchy of equations:
\begin{align}
-2\partial_+\partial_-\tilde{h}^{(1,0)}_{ij} + \Box_\perp\tilde{h}^{(1,0)}_{ij} &= \kappa^2 S^{(1,0)}_{ij}\,, \label{eq:h10}\\
\bar{g}_{--}\partial_+^2\tilde{h}^{(1,0)}_{ij} - 2\partial_+\partial_-\tilde{h}^{(1,1)}_{ij} + \Box_\perp\tilde{h}^{(1,1)}_{ij} &= \kappa^2 S^{(1,1)}_{ij}\,, \label{eq:h11}\\
\bar{g}_{--}\partial_+^2\tilde{h}^{(1,1)}_{ij} - 2\partial_+\partial_-\tilde{h}^{(1,2)}_{ij} + \Box_\perp\tilde{h}^{(1,2)}_{ij} &= \kappa^2 S^{(1,2)}_{ij}\,. \label{eq:h12}
\end{align}
This hierarchy structure holds for the dilute-dense approximation to all orders. Each order in $\mu_H$ is sourced by the solution at the previous order, multiplied by the background metric $\bar{g}_{--}$. 

In the remainder of this section, we will rederive our previous results in \cite{Raj:2023iqn} for order $O(\mu_L)$ and $O(\mu_L \mu_H)$. We first expand $\Delta$ in Eq.~\eqref{Delta-def} in powers of $\mu_H$ using Eq.~\eqref{x-tilde-shorthand}. It will be useful to introduce the shorthand notation
\begin{align}
\label{eq:short-hand}
F(\boldsymbol{x}) \equiv \frac{\rho_H(\boldsymbol{x})}{\Box_\perp},
\qquad
A_i(\boldsymbol{x}) \equiv \partial_i F(\boldsymbol{x}) .
\end{align}
We first expand $\Delta$ for a point particle at fixed transverse
position $\boldsymbol{b}$.  Defining
\begin{align}
\Delta_0^{(\boldsymbol b)}
\equiv
\delta(x^+)\delta(x^-)
\delta^{(2)}(\boldsymbol{x}-\boldsymbol{b}),
\end{align}
Eqs.~\eqref{Delta-def} and \eqref{x-tilde-shorthand} give
\begin{align}
\Delta
={}&
\delta(x^+)\delta^{(2)}(\boldsymbol{x}-\boldsymbol b)+\kappa^2\mu_H
\left[
F(\boldsymbol b)\frac{\partial_+}{\partial_-}
+
A_k(\boldsymbol b)
\frac{\partial_k}{\partial_-^2}
\right]
\Delta_0^{(\boldsymbol b)}
\nonumber\\
&+\kappa^4\mu_H^2
\Bigg[
-\frac12 A_k(\boldsymbol b)A_k(\boldsymbol b)
\frac{\partial_+}{\partial_-^2}
+\frac12 F^2(\boldsymbol b)
\frac{\partial_+^2}{\partial_-}
\nonumber\\
&\hspace{1.7cm}
+A_k(\boldsymbol b)A_l(\boldsymbol b)
\frac{\partial_k\partial_l}{\partial_-^3}
+
F(\boldsymbol b)A_k(\boldsymbol b)
\frac{\partial_+\partial_k}{\partial_-^2}
\Bigg]
\Delta_0^{(\boldsymbol b)}
+\mathcal O(\mu_H^3).
\label{eq:Delta_expand}
\end{align}
In the above formula, the smearing from point particles to a transverse density profile $\rho_L(\boldsymbol{x})$,
\begin{equation}
\label{smear}
\Delta_0 \to \rho_L(\boldsymbol{x})\,\delta(x^+)\delta(x^-)\,,
\end{equation}
must be performed before replacing $\boldsymbol b$ by
$\boldsymbol x$.  Specifically,
\begin{align}
\Delta[\rho_L](x)
\equiv
\int d^2\boldsymbol b\,
\rho_L(\boldsymbol b)\,
\Delta^{(\boldsymbol b)}(x).
\label{eq:smear-definition}
\end{align}
For terms containing derivatives of the transverse delta function, the correct prescription is
\begin{align}
&\int d^2\boldsymbol b\,
\rho_L(\boldsymbol b)\,
{\cal A}(\boldsymbol b)\,
\partial_{i_1}\cdots\partial_{i_n}
\delta^{(2)}(\boldsymbol x-\boldsymbol b)=
\partial_{i_1}\cdots\partial_{i_n}
\left[
{\cal A}(\boldsymbol x)\rho_L(\boldsymbol x)
\right].
\label{eq:smear-identity}
\end{align}
The correctly smeared expansion is
\begin{align}
\label{eq:Delta-expand-smeared}
\Delta[\rho_L]
={}&
\Delta_0
+\kappa^2\mu_H
\left[
F\frac{\partial_+}{\partial_-}\Delta_0
+
\frac{1}{\partial_-^2}
\partial_k\left(A_k\Delta_0\right)
\right]
\\
+&\kappa^4\mu_H^2
\Bigg[
-\frac12 A_kA_k
\frac{\partial_+}{\partial_-^2}\Delta_0
+\frac12F^2
\frac{\partial_+^2}{\partial_-}\Delta_0
+
\frac{1}{\partial_-^3}
\partial_k\partial_l
\left(A_kA_l\Delta_0\right)
+
\frac{\partial_+}{\partial_-^2}
\partial_k\left(FA_k\Delta_0\right)
\Bigg]
+\mathcal O(\mu_H^3)~.\nonumber
\end{align}

\subsubsection{Solution at order $\mu_L$}
\label{sec:zeroth-order}
The leading-order equation in Eq.\eqref{eq:h10} reduces to the flat-space wave equation:
\begin{equation}
-\Box\tilde{h}^{(1,0)}_{ij} = \kappa^2\mu_L(2\partial_i\partial_j - \Box_\perp\delta_{ij})\frac{1}{\partial_+^2}\(\delta(x^+)\delta(\bsx-\bsb) \)\,,
\end{equation}
where $\Box = 2\partial_+\partial_- - \Box_\perp$ is the flat-space d'Alembertian. Using $\delta(x^+)/\partial_+^2 = x^+\Theta(x^+)$ and inverting $\Box_\perp$, we obtain
\begin{equation}
\tilde{h}^{(1,0)}_{ij} = \kappa^2\mu_L\,x^+\Theta(x^+)\left(\frac{2\partial_i\partial_j}{\Box_\perp} - \delta_{ij}\right)\delta^{(2)}(\boldsymbol{x}-\boldsymbol{b})\,.
\label{eq:h10_sol}
\end{equation}
This result is nothing but the perturbative Aichelburg-Sexl solution in lightcone gauge $h_{+\mu}=0$ for the gravitational wave generated by the light source.

\subsubsection{Solution at order $\mu_L\mu_H$}
\label{sec:Lipatov-vertex-calculation}
At order $\mu_L\mu_H$, the correctly smeared stress tensor gives
\begin{align}
T_{++}^{(1,1)}
&=
\kappa^2\mu_L\mu_H
\left[
F\frac{\partial_+}{\partial_-}\Delta_0
+
\frac{1}{\partial_-^2}
\partial_k\left(A_k\Delta_0\right)
\right]=\kappa^2\mu_L\mu_H
\left[
F\frac{\partial_+}{\partial_-}
+
A_k\frac{\partial_k}{\partial_-^2}
+
\rho_H\frac{1}{\partial_-^2}
\right]\Delta_0 ,
\label{eq:Tpp11}
\\[5pt]
T_{+i}^{(1,1)}
&=
\kappa^2\mu_L\mu_H
A_i\frac{\Delta_0}{\partial_-},
\qquad
T_{ij}^{(1,1)}=T^{(1,1)}=0 .
\end{align}
In the second equality in Eq.~\eqref{eq:Tpp11}, we used
$\partial_kA_k=\Box_\perp F=\rho_H$. The last term in the expression for $T_{++}^{(1,1)}$ is the contact term of \cite{Raj:2023irr}. As we discussed there at length, this term is crucial to obtain the $N^\mu N^\nu$ term, which appears the gravitational Lipatov vertex (see Eq.~\eqref{eq:lipatov_dc}), and was argued by Lipatov to be necessary to preserve unitarity~\cite{Lipatov:1982it}. Here, using the smearing prescription given in Eqs.~\eqref{smear}, \eqref{eq:smear-definition} and \eqref{eq:smear-identity}, we have clarified that the origin of the term follows from the proper handling of the derivatives of the delta function.

Going through each term in $S_{ij}$ and using Eq.~\eqref{eq:short-hand} we find
\begin{align}
S_{ij}^{(1,1)} &= \mu_L\mu_H\kappa^2\bigg\{(2\partial_i\partial_j - \Box_\perp\delta_{ij})\left[\frac{\rho_H(\boldsymbol{x})}{\Box_\perp}\frac{\Delta_0}{\partial_-\partial_+} + \partial_k\(\frac{\partial_k\rho_H(\boldsymbol{x})}{\Box_\perp}\frac{1}{\partial_-^2\partial_+^2}\Delta_0\)\right] - \no\\[15pt]
&\frac{2}{\partial_+\partial_-}\left[\partial_i\left(\frac{\partial_j\rho_H(\boldsymbol{x})}{\Box_\perp}\Delta_0\right) + \partial_j\left(\frac{\partial_i\rho_H(\boldsymbol{x})}{\Box_\perp}\Delta_0\right) - \delta_{ij}\partial_k\left(\frac{\partial_k\rho_H(\boldsymbol{x})}{\Box_\perp}\Delta_0\right)\right]\bigg\}\,.
\end{align}
Next, moving $\bar{g}_{--}\partial_+^2\tilde{h}^{(1,0)}_{ij}$ in Eq.~\eqref{eq:h11} to the r.h.s, we obtain 
\begin{align}
-\square \tilde{h}_{ij}^{(1,1)} &= \mu_L\mu_H\kappa^4\bigg\{(2\partial_i\partial_j - \Box_\perp\delta_{ij})\left[\frac{\rho_H(\boldsymbol{x})}{\Box_\perp}\frac{\Delta_0}{\partial_-\partial_+} + \partial_k\(\frac{\partial_k\rho_H(\boldsymbol{x})}{\Box_\perp}\frac{1}{\partial_-^2\partial_+^2}\Delta_0\)\right]\no\\[15pt]
&- \frac{2\rho_H(\boldsymbol{x})}{\Box_\perp}\left(\frac{2\partial_i\partial_j}{\Box_\perp} - \delta_{ij}\right)\Delta_0  \no\\[15pt]
&-\frac{2}{\partial_+\partial_-}\left[\partial_i\left(\frac{\partial_j\rho_H(\boldsymbol{x})}{\Box_\perp}\Delta_0\right) + \partial_j\left(\frac{\partial_i\rho_H(\boldsymbol{x})}{\Box_\perp}\Delta_0\right) - \delta_{ij}\partial_k\left(\frac{\partial_k\rho_H(\boldsymbol{x})}{\Box_\perp}\Delta_0\right)\right]\bigg\}\,.
\end{align}
Performing the smearing operation in Eq.~\eqref{smear}, taking the Fourier transform\footnote{The convention of the FT is
\begin{align}
    f(x) = \int \frac{d^4q}{\(2\pi\)^4}~ e^{iq^+x^-+iq^-x^+-i\bsq\cdot \bsx} \tilde{f}(q)~,\qquad \tilde{f}(q) = \int d^4x~ e^{-iq^+x^--iq^-x^++i\bsq\cdot \bsx} f(x)
\end{align}} of this expression, and setting the 4-momentum of the outgoing graviton to zero ($2k^+k^- = \bsk^2$), gives the final dilute-dilute result 
\begin{align}
\label{lipatov-vertex}
     \hat{h}_{ij}^{(1,1)}(k) = \frac{2\mu_L\mu_H\kappa^4}{k^2+i\epsilon k^-} \int \frac{d\bsq^2_2}{(2\pi)^2} \Gamma_{ij}(\bsq_1, \bsq_2) \frac{\rho_H(\bsq_1)}{\bsq_1^2}\frac{\rho_L(\bsq_2)}{\bsq_2^2}\,,
\end{align}
where $\Gamma_{ij}$ is the expression for the gravitation Lipatov vertex in  lightcone gauge \cite{Raj:2023irr}
\begin{align}
\label{Lipatov-LC}
    \Gamma_{ij}(\boldsymbol{q}_1, \boldsymbol{q}_2) = 2\left(q_{2i} - k_i\frac{\boldsymbol{q}_2^2}{\boldsymbol{k}^2}\right)
\left(q_{2j} - k_j\frac{\boldsymbol{q}_2^2}{\boldsymbol{k}^2}\right)
- 2k_ik_j\frac{\boldsymbol{q}_1^2\boldsymbol{q}_2^2}{\boldsymbol{k}^4}\,,
\end{align}
with $\bsk = \bsq_1+\bsq_2$.

\subsection{Angular structure of Lipatov radiation at $O(\mu_L\mu_H)$}
\label{subsec:angular}
Before we consider the next $O(\mu_L\mu_H^2)$ corrections, we will digress to consider the angular structure of gravitational radiation from the Lipatov vertex in lightcone gauge.
This was previously analyzed by Ciafaloni, Colferai Veneziano (CCV) and Coradeschi
(CCCV)~\cite{Ciafaloni:2015xsr,Ciafaloni:2015vsa} within the context of multiple scattering contributions in the soft (Weinberg) and semi-hard (Lipatov) radiation regimes. As shown by CCCV, and expanded upon in later work in the ACV framework~\cite{Gruzinov:2014moa,Ciafaloni:2016nul,Ciafaloni:2017ort,Ciafaloni:2018uwe},  these contributions qualitatively modify the large frequency spectrum of gravitational radiation away from the weak dependence seen in the low-frequency Smarr limit~\cite{Smarr:1977fy}. 

The CCCV derivation is interesting in its own right and is reproduced in  Appendix~\ref{app:CCV}. It involves careful matching of the Weinberg and Lipatov regimes in their formalism. It is not however obvious that this matching is identical in the shockwave framework. To understand this better, we will first work out in detail the angular structure of Lipatov radiation. 
We will then show that in the soft limit, this angular structure is identical to that of the {\it ultrarelativistic limit} of that obtained from the Weinberg current. The consequences of this matching vis-\`{a}-vis CCCV will be discussed at the end of this subsection. 

\begin{figure}[ht]
\centering
\includegraphics[scale=0.8]{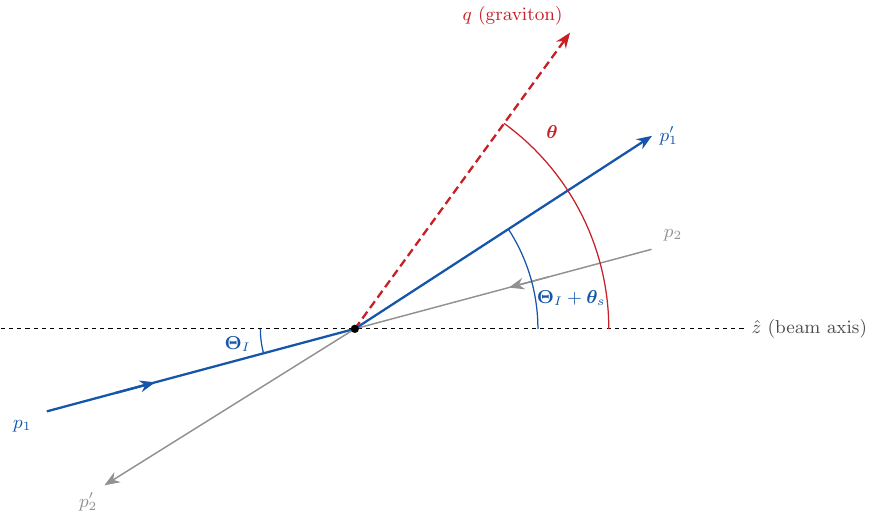}
\caption{Description of the angles relevant for Lipatov radiation. The collision beam axis $\hat z$ lies along the horizontal axis. The incoming momentum $p_1$ has a small incidence angle $\Theta_I$ with $\hat z$. After the scattering, it emerges as $p'_1$ at angle
$\Theta_I+\bm{\theta}_s$. The radiated graviton $q$ is emitted at angle
$\bm{\theta}$. Each of these angles is a vector in the 2-D plane of the scattering.}
\label{fig:angles}
\end{figure}

We begin by considering the emission of a soft graviton accompanying the elastic
scattering of two ultrarelativistic particles,
\begin{equation}
p_1 + p_2 \;\longrightarrow\; p'_1 + p'_2 \;+\; \text{graviton}(k).
\end{equation}
Fig.~\ref{fig:angles} provides a visual dictionary of the relevant angles in the center-of mass frame. In eikonal kinematics, all these angles are small:  $|\bth|,\,|\bTh_I|,\,|\bth_s|\ll 1$.  The massless momenta can be parametrized in terms of these angles as follows:
\begin{align}
\label{eq:mom-1}
p_1   &= E\,(1,\,\bTh_I,\,\sqrt{1-\Th_I^2}), \qquad 
p'_1  = E\,(1,\,\bTh_I+\bth_s,\,\sqrt{1-|\bTh_I+\bth_s|^2}),
\end{align}
with $p_2,\,p'_2$ being nearly anti-parallel to $\hat z$ (the collision beam axis). The 2-vector $\bTh_I$ is the \emph{incidence angle} (as indicated by the subscript) of $p_1$, namely, the
direction $p_1$ makes with $\hat z$. The 2-vector $\bth_s$ is the small
deflection angle ($p'_1$ is obtained from $p_1$ by adding $\bth_s$). The graviton's 4-momentum is similarly parametrized as
\begin{equation}
\label{eq:mom-2}
k^\mu = \omega\,(1,\,\bth,\,\sqrt{1-\theta^2}),\qquad |\bth|\ll 1,
\end{equation}
The 2-vector $\bth$ is the graviton emission angle (relative to $\hat z$). We can write its modulus and azimuth as $|\bth|=\theta$ and $\fp_\theta$; in the complex 2-vector notation\footnote{Some simple useful identities in the complex number ``tilde" notation ${\tilde v}\equiv v_x+iv_y$ for the transverse 2--vector $\bm v=(v_x,v_y)$ are 
\begin{equation}
\label{eq:complex-vec}
|v|^2 = {\tilde v}\,{\tilde v}^* = v_x^2+v_y^2, \qquad
\fp_{v} = \arg {\tilde v}, \qquad
\frac{{\tilde v}^2}{|v|^2} = e^{2i\fp_v}.
\end{equation}} this reads as ${\tilde \theta}\equiv\theta_x+i\theta_y = |\theta|\,e^{i\fp_\theta}$.

In this complex notation, the QCD Lipatov vertex $C_i \equiv q_{2i} - k_i\bsq_2^2/\bsk^2$ can be written as 
\be
\label{app-Ctilde}
\tilde{C} = \tilde{q}_2 - \tilde{k}\,\frac{\tilde{q}_2\tilde{q}_2^*}{\tilde{k}\tilde{k}^*} = \tilde{q}_2\,\frac{\tilde{k}^*-\tilde{q}_2^*}{\tilde{k}^*} = \frac{\tilde{q}_2\,\tilde{q}_1^*}{\tilde{k}^*}\,,
\ee
where in the last step we used momentum conservation $\tilde{k}-\tilde{q}_2=\tilde{q}_1$. With this form, the contraction of the gravitational Lipatov vertex in Eq.~\eqref{Lipatov-LC} with the negative-helicity polarization tensor $\eps^{(-2)}_{ij}$ can be expressed as 
\be
\label{eq:temp-1}
\eps^{(-2)}_{ij}\Gamma_{ij} = (\tilde{C}^*)^2 - \frac{(\tilde{k}^*)^2\,\bsq_1^2\bsq_2^2}{\bsk^4} = \frac{(\tilde{q}_2^*)^2\,\tilde{q}_1^2 - \bsq_1^2\,\bsq_2^2}{\tilde{k}^2}\,.
\ee
where\footnote{More explicitly,
\begin{align}
\epsilon_{\mu\nu}^{(-2 )}=\cfrac{1} {2} \left( \begin{matrix} {0} & {0} & {0} & {0} \\ {0} & {1} & {-i} & {0} \\ {0} & {-i} & {-1} & {0} \\ {0} & {0} & {0} & {0} \\ \end{matrix} \right).
\end{align}.
Note further that the second equality in Eq.~\eqref{eq:temp-1} uses $(\tilde{k}^*)^2/\bsk^4=1/\tilde{k}^2$ and Eq.~\eqref{app-Ctilde}.}
\begin{align}
\label{polarization-tensor}
    \epsilon_{\mu\nu}^{(-2 )}:=\epsilon_{\mu}^{(-)} \epsilon_{\nu}^{(-)} ,\qquad \epsilon^{(-)}_{\mu} = \frac{1}{\sqrt{2}}(0, 1, -i, 0).
\end{align}

Writing each momentum in polar form so that $(\tilde{q}_2^*)^2\tilde{q}_1^2 = \bsq_1^2\bsq_2^2\,e^{2i(\f_{q_1}-\f_{q_2})}$, and using $1/\tilde{k}^2=e^{-2i\f_k}/\bsk^2$, this expression can be further simplified to read 
\be
\label{app-Gam-proj}
\eps^{(-2)}_{ij}\,\Gamma_{ij}(\bsq_1,\bsq_2) = -\,\frac{\bsq_1^2\,\bsq_2^2}{\bsk^2}\,e^{-2i\f_k}\,\Bigl[1-e^{-2i(\f_{q_2}-\f_{q_1})}\Bigr]\,.
\ee
The factors $\bsq_1^2$ and $\bsq_2^2$ in the numerator exactly cancel the $t$-channel propagators $1/\bsq_1^2$ and  $1/\bsq_2^2$ in Eq.~\eqref{lipatov-vertex}, leaving an integrand expressed entirely in terms of the graviton transverse momentum scale $|\bsk|$ and the helicity phase difference $\f_{q_2}-\f_{q_1}$ between the two classical (reggeons) fields.

Substituting Eq.~\eqref{app-Gam-proj} into Eq.~\eqref{lipatov-vertex}, taking the on-shell residue at $k^2=0$ and using the Fourier transforms $\rho_L(\bsx)=\delta^{(2)}(\bsx-\bsb) \implies \rho_L(\bsq_2)=e^{-i\bsq_2\cdot\bsb}$ and $\rho_H(\bsx)=\delta^{(2)}(\bsx) \implies \rho_H(\bsq_1)= 1$, the {\it on-shell} emission amplitude for single graviton exchange is 
\begin{align}
\mathcal M_{\rm Lipatov}^{(1,1)}(b,E;\omega,\boldsymbol{\theta})
&\equiv
\lim_{k^2\to0}
\left(
k^2+i\epsilon k^-
\right)
\frac{1}{\kappa}\epsilon^{(-2)}_{ij}
\tilde h_{ij}^{(1,1)}(k)~,\nonumber\\[5pt]
&=s\kappa^3\,
\int\frac{d^2\bsq_2}{(2\pi)^2\,\bsk^2}\,e^{-i\bsq_2\cdot\bsb}\,e^{-2i\f_k}\,
\bigg[1-\exp\left(-2i(\f_{q_2}-\f_{q_2-k})\right)\bigg]\,,
\label{app-M1-b}
\end{align}
where $\bsq_2= E\boldsymbol{\theta}_s$ is the elastic momentum transfer from the reggeon emitted by the light source in the expression of $\Gamma_{ij}$ and $\bsk$ is the graviton transverse momentum. Recall that the graviton transverse momentum is parametrized as $\bsk=\omega\boldsymbol{\theta}$, where $\boldsymbol{\theta}$ is the angle of graviton emission in the transverse 2D space.  The phase difference within the parenthesis is the helicity-2 phase transfer between the two reggeons $\bsq_2$ and $\bsq_2-\bsk=-\bsq_1$ at the Lipatov vertex. It is evident that at large $\omega$ the amplitude $\mathcal M_{\rm Lipatov}^{(1,1)} \sim 1/\omega^2$. Hence, the spectrum from this order is 
\begin{equation}
\label{eq:LO-large-omega}
\frac{dE_{GW}}{d\omega}|_{\rm LO}
\sim \omega^2\,\bigl|\mathcal M_{\rm Lipatov}^{(1,1)}\bigr|^2\sim \omega^{-2}. 
\end{equation}

On the other hand, for small $\omega$ we should expect the universal $1/\omega$ behaviour that we demonstrate next \cite{Weinberg:1965nx}.

\begin{figure}[ht]
\centering
$$
\vcenter{\hbox{
    \includegraphics[scale=0.55]{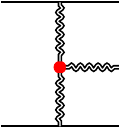}
}}
\;=\;
\vcenter{\hbox{
    \includegraphics[scale=0.55]{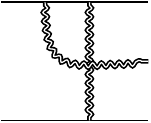}
}}
\;+\;
\vcenter{\hbox{
    \includegraphics[scale=0.55]{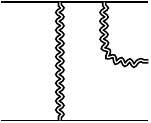}
}}
\;+\;
\vcenter{\hbox{
    \includegraphics[scale=0.55]{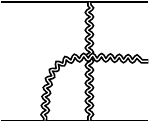}
}}
\;+\;
\vcenter{\hbox{
    \includegraphics[scale=0.55]{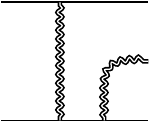}
}}
\;+\;
\vcenter{\hbox{
    \includegraphics[scale=0.55]{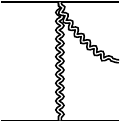}
}}
\;+\;
\vcenter{\hbox{
    \includegraphics[scale=0.55]{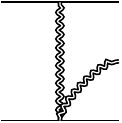}
}}
\;+\;
\vcenter{\hbox{
    \includegraphics[scale=0.55]{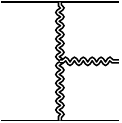}
}}
$$
\caption{Decomposition of the gravitational Lipatov vertex (denoted by the red dot) in terms of bare Feynman diagrams.}
\label{fig:effective-vertex-decomposition-2}
\end{figure}
\begin{figure}[ht]
    \centering
    \includegraphics[scale=0.75]{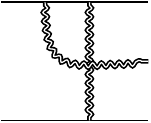}\qquad
    \includegraphics[scale=0.75]{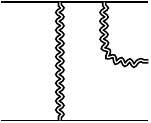}\qquad
    \includegraphics[scale=0.75]{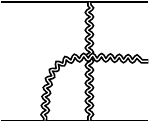}\qquad
    \includegraphics[scale=0.75]{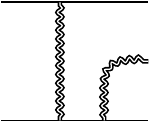}
    \caption{Gravitational emission diagrams that contribute to the Weinberg current. These diagrams are enhanced by $1/\omega$, originating from the eikonal propagator of the external lines.}
    \label{fig:W1-W4}
\end{figure}
\begin{figure}[ht]
    \centering
    \includegraphics[scale=0.75]{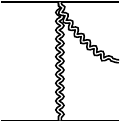}\qquad
    \includegraphics[scale=0.75]{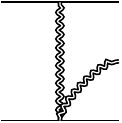}\qquad
    \includegraphics[scale=0.75]{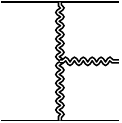}
    \caption{Gravitational emission diagrams  that are $O(\omega^0)$ for small $\omega$. These diagrams become important as one moves away from the Weinberg regime.}
    \label{fig:W5-W7}
\end{figure}

In GR, the Lipatov vertex, in multi-Regge kinematics can be written as the sum of the diagrams depicted in Fig.~\ref{fig:effective-vertex-decomposition-2}~\cite{SabioVera:2011wy}. However, for $\omega\rightarrow 0$, the emissions from the incoming and outgoing lines shown in Fig.~\ref{fig:W1-W4} are dominant and comprise the Weinberg current. The graviton self-interaction diagrams shown in Fig.~\ref{fig:W5-W7} are $O(\omega^0)$ and only begin to contribute beyond the soft limit. As shown explicitly in \cite{Raj:2025hse}, the soft limit of the Lipatov current in Fig.~\ref{fig:effective-vertex-decomposition-2} equals the ultrarelativistic limit of the Lipatov current. 

We will follow a different tack here relative to \cite{Raj:2025hse} and use the angular structure of the Lipatov amplitude to recover the ultrarelativistic limit of the Weinberg result. It is sufficient for this purpose to examine the integrand of Eq.~\eqref{app-M1-b}, which we define to be 
\begin{align}
\mathcal{I}_{\text{L}} = \frac{e^{-2i\f_k}}{\bsk^2}\bigg[1-\exp\left(-2i(\f_{q_2}-\f_{q_2-k})\right) \bigg]
\end{align}
To analyze the soft limit of this expression, it is convenient to rewrite it as follows
\begin{equation}
\mathcal{I}_{\text{L}} \;=\; \frac{e^{-2i\f_k}}{\bsk^2}\left(1 - \frac{\tilde q_2^*}{\tilde q_2}\frac{\tilde q_2-\tilde k}{(\tilde q_2-\tilde k)^*}\right) \;=\; \frac{e^{-2i\f_k}}{\bsk^2}\;\frac{\tilde k \tilde q_2^{*}-\tilde k^{*}\tilde q_2}{\tilde q_2\,(\tilde q_2-\tilde k)^{*}} \;=\; \frac{e^{-2i\f_k}}{\omega E}\frac{1}{|\bth^2|} \left( \frac{\tilde \theta\tilde \theta_s^{*} - \tilde \theta^{*}\tilde \theta_s}{\tilde \theta_s(\tilde \theta_s- x\tilde \theta)^{*}}\right)~,
\label{eq:IRegge}
\end{equation}
where we defined $x=\omega/E$. In this form, it is trivial to see that the soft limit, where $k\to 0$ (or equivalently $x\to 0$) is
\begin{equation}
\mathcal{I}_{\text{L}}\Big|_{\rm soft} \;=\;  \frac{e^{-2i\f_k}}{\omega E}\frac{\tilde \theta\tilde \theta_s^{*} - \tilde \theta^{*}\tilde \theta_s}{|\bth^2||\bth^2_s|}~.
\label{eq:IRegge-1}
\end{equation}

We will now show that this expression agrees with the high energy limit of the Weinberg soft amplitude. We begin with the high energy limit of the Weinberg amplitude derived in \cite{Raj:2025hse}, and write it in the impact parameter representation as 
\begin{equation}
\mathcal{M}_{\text{Weinberg}}\Big|_{\rm HE} \;=\; s\kappa^3 \int\!\frac{d^2\qv}{(2\pi)^2}\;
e^{i\qv\cdot\bv}\;\mathcal{I}_{\text{W}}(\kv,\qv)\Big|_{\rm HE} \ ,
\label{eq:m_soft}
\end{equation}
where\footnote{The vertex in Eq.~(3.103) of \cite{Raj:2025hse} is a pure emission vertex. The full amplitude carries in addition the elastic amplitude 
$\mathcal{M}_{\rm el}\propto \frac{1}{t}= -\,\frac{1}{\qv^2}$.
After appropriate normalization, the integrand of the Weinberg amplitude takes the form shown here.}
\begin{align}
    \mathcal{I}_{\text{W}}(\kv,\qv)\Big|_{\rm HE} = \frac{4}{\bsq_2^2}\left[-\frac{\bsk\cdot \bsq_1}{\bm{k}^4}\,k^ik^j
        + \frac{k^i q_1^j}{\bm{k}^2}\right]\epsilon_{ij}^{(-)} \ , \qquad \bsk  = \bsq_1 + \bsq_2~.
\end{align}
For the projection of the terms in the bracket to the negative helicity polarization tensor, we use the identities
\begin{equation}
k^ik^j\epsilon^{(-)}_{ij} = \tfrac12 (\tilde k^{*})^2 \ ,
\qquad
k^i q_1^j\epsilon^{(-)}_{ij} = \tfrac12 \tilde k^{*} \tilde q_1^{*} \ ,
\qquad
\bm{k}\!\cdot\!\bm{q}_1 = \tfrac12\big(\tilde k \tilde q_1^{*} + \tilde k^{*}\tilde q_1\big) \ .
\label{mtc:contractions}
\end{equation}
This simplifies the expression to
\begin{align}
    \mathcal{I}_{\text{W}}(\kv,\qv)\Big|_{\rm HE} = -\frac{\tilde k^* \tilde q_1 - \tilde k \tilde q_1^*}{\bsq_2^2 \tilde{k}^2} =  \frac{e^{-2i\f_k}}{\omega E}\frac{\tilde \theta\tilde \theta_s^{*} - \tilde \theta^{*}\tilde \theta_s}{|\bth^2||\bth^2_s|}~.
    \label{eq:I_soft-HE}
\end{align}
From Eqs.~\eqref{eq:IRegge-1} and \eqref{eq:I_soft-HE}, we see that the respective limits of the Lipatov and the Weinberg integrands agree:
\begin{align}
    \mathcal{I}_{\text{L}}\Big|_{\rm soft} = \mathcal{I}_{\text{W}}\Big|_{\rm HE}~.
\end{align}
The transverse momentum integral in Eq.~\eqref{eq:m_soft} can  be performed directly using complex coordinates in the $\boldsymbol{\theta}_s$ plane. After some algebra, we 
obtain
\begin{align}
\mathcal{M}_{\rm Lipatov}^{(1,1)}\Big|_{\rm \omega\rightarrow 0}
=
-\frac{s\kappa^3}{\pi}\,e^{-2i\phi_k}\frac{\bsk \times \bsb}{|\bsk|^2 |\bsb|^2}\,.
\label{eq:Lipatov-soft-b-space}
\end{align} 
This result shows that the soft limit of the Lipatov radiation amplitude vanishes identically when the graviton emission is in the direction of the impact parameter.  
This is precisely what one observes for a sheet of plane polarized radiation {\it a la} that of the Weizs\"{a}cker-Williams fields of an ultrarelativistic electromagnetically charged object. 

\section{$O(\mu_L\mu_H^2)$ tidal contribution to Lipatov radiation}
\label{sec:NLO}
In this section, we will compute the first rescattering correction to gravitational wave radiation going beyond the dilute-dilute limit. For this, we will need to solve Eq.~\eqref{eq:h12}; as a first step, we need to calculate $S_{ij}^{(1,2)}$ and take its Fourier transform.  We will subsequently compute the Fourier transform of the contribution $\bar{g}_{--}\partial_+^2\tilde{h}^{(1,1)}_{ij} $ from the previous order that contributes to Eq.~\eqref{eq:h12}.

\subsection{Contribution from source terms}
\label{sec:source-terms}
To calculate $S_{ij}^{(1,2)}$, we first collect the components of the
energy-momentum tensor at order $O(\mu_L\mu_H^2)$.  Using the  smeared expansion in Eq.~\eqref{eq:Delta-expand-smeared}, we obtain
\begin{align}
T_{++}^{(1,2)}
={}&
\kappa^4\mu_L\mu_H^2
\Bigg[
-\frac12 A_kA_k
\frac{\partial_+}{\partial_-^2}\Delta_0
+\frac12F^2
\frac{\partial_+^2}{\partial_-}\Delta_0
+
\frac{1}{\partial_-^3}
\partial_k\partial_l
\left(A_kA_l\Delta_0\right)
+
\frac{\partial_+}{\partial_-^2}
\partial_k\left(FA_k\Delta_0\right)
\Bigg],
\label{eq:Tpp12}
\\[10pt]
T_{+i}^{(1,2)}
={}&
\kappa^4\mu_L\mu_H^2
\left[
FA_i\frac{\partial_+}{\partial_-}\Delta_0
+
\frac{1}{\partial_-^2}
\partial_k\left(A_iA_k\Delta_0\right)
\right]
\equiv{}
\kappa^4\mu_L\mu_H^2\,{\cal P}_i(x),
\label{eq:Tpi12}
\\[10pt]
T_{ij}^{(1,2)}
={}&
\kappa^4\mu_L\mu_H^2
A_iA_j\frac{1}{\partial_-}\Delta_0 .
\label{eq:Tij12}
\end{align}
Next, substituting Eqs.~\eqref{eq:Tpp12}--\eqref{eq:Tij12} into the
definition of $S_{ij}$ in Eq.~\eqref{eq:Sij} gives
\begin{align}
S_{ij}^{(1,2)}
={}&
\kappa^4\mu_L\mu_H^2
\Bigg\{
(2\partial_i\partial_j-\Box_\perp\delta_{ij})
\Bigg[
-\frac12A_kA_k
\frac{1}{\partial_-^2\partial_+}\Delta_0
+\frac12F^2\frac{1}{\partial_-}\Delta_0
+\frac{1}{\partial_-^3\partial_+^2}
\partial_k\partial_l
\left(A_kA_l\Delta_0\right)
\nonumber\\
+&
\frac{1}{\partial_-^2\partial_+}
\partial_k\left(FA_k\Delta_0\right)
\Bigg]+
\left(
2A_iA_j-\delta_{ij}A_kA_k
\right)
\frac{\Delta_0}{\partial_-}
-
\frac{2}{\partial_+}
\left[
\partial_i{\cal P}_j
+\partial_j{\cal P}_i
-\delta_{ij}\partial_k{\cal P}_k
\right]
\Bigg\}.
\label{source12}
\end{align}

We will now compute the Fourier transform ${\tilde S}_{ij}$ of the above expression. Pulling out the common integral and factors $\int\frac{d^2\boldsymbol{q}_1d^2\boldsymbol{q}_1'd^2\boldsymbol{q}_2}{(2\pi)^4}\delta^{(2)}(\boldsymbol{k}-\boldsymbol{q}_1-\boldsymbol{q}_1'-\boldsymbol{q}_2)\frac{\tilde{\rho}_H(\boldsymbol{q}_1)}{\boldsymbol{q}_1^2}\frac{\tilde{\rho}_H(\boldsymbol{q}_1')}{\boldsymbol{q}_1'^2}\tilde{\rho}_L(\boldsymbol{q}_2)$, the terms with the structure 
$(2\partial_i\partial_j-\Box_\perp\delta_{ij})[\cdots]$ give 
\begin{align}
\label{line1}
(-2k_i k_j+\bsk^2\delta_{ij})
\Bigg[
&\frac{1}{2ik^+}
+\frac{i(\bsq_1\cdot\bsq_1')}
       {2(k^+)^2k^-}
+\frac{
(\bsq_1\cdot\bsk)(\bsq_1'\cdot\bsk)
}{
i(k^+)^3(k^-)^2
}
-
\frac{
i(\bsq_1+\bsq_1')\cdot\bsk
}{
2(k^+)^2k^-
}
\Bigg].
\end{align}
Next, the terms $\left(
2A_iA_j-\delta_{ij}A_kA_k
\right)[\cdots]$ give
\begin{align}
\label{line2}
-\frac{1}{ik^+}
\left(
q_{1i}q_{1j}'
+q_{1i}'q_{1j}
-\delta_{ij}\bsq_1\cdot\bsq_1'
\right)\,.
\end{align}
To compute the Fourier transform of $\frac{2}{\partial_+}
\left[
\partial_i{\cal P}_j
+\partial_j{\cal P}_i
-\delta_{ij}\partial_k{\cal P}_k
\right]$, we first define 
\begin{align}
\boldsymbol{s} &\equiv \boldsymbol{q}_1+\boldsymbol{q}_1'~,\qquad
Y_i \equiv
q_{1i}(\boldsymbol{q}_1'\cdot\boldsymbol{k})
+
q_{1i}'(\boldsymbol{q}_1\cdot\boldsymbol{k}) .
\end{align}
In these variables, we obtain 
\begin{align}
\label{line3}
\frac{1}{ik^-}
\left[
k_iX_j+k_jX_i-\delta_{ij}\bsk\cdot\boldsymbol X
\right]~,\qquad \text{where}~
X_i
\equiv
\frac{k^-}{k^+}s_i
+
\frac{Y_i}{(k^+)^2}.
\end{align}
Combining Eqs.~\eqref{line1}--\eqref{line3}, and imposing the on-shell condition $2k^+k^-=\bsk^2$, the result can be written as $\tilde S_{ij}=\widetilde S_{ij}(\bsq_1,\bsq_1',\bsq_2)/(ik^+)$, with 
\begin{align}
\label{Sij-FT}
\widetilde S_{ij}
(\bsq_1,\bsq_1',\bsq_2)
=\alpha
\left(
-2k_i k_j+\bsk^2\delta_{ij}
\right)
-
\left(
q_{1i}q_{1j}'
+q_{1i}'q_{1j}
-\delta_{ij}\bsq_1\cdot\bsq_1'
\right)
+
\left(
k_i r_j+k_j r_i
-\delta_{ij}\bsk\cdot\boldsymbol r
\right),
\end{align}
where 
\begin{align}
\alpha
=&
\frac12
+
\frac{
\boldsymbol{s}\cdot\boldsymbol{k}
-\boldsymbol{q}_1\cdot\boldsymbol{q}_1'
}{
\bsk^2
}
+
\frac{
4(\boldsymbol{q}_1\cdot\boldsymbol{k})
 (\boldsymbol{q}_1'\cdot\boldsymbol{k})
}{
\bsk^4
},
\label{eq:alpha-corrected}
\\
\boldsymbol r
=&
\boldsymbol s
+
\frac{2}{\bsk^2}
\left[
\boldsymbol q_1
(\boldsymbol q_1'\cdot\boldsymbol k)
+
\boldsymbol q_1'
(\boldsymbol q_1\cdot\boldsymbol k)
\right],
\qquad
\boldsymbol{s}
\equiv
\boldsymbol q_1+\boldsymbol q_1'.
\label{eq:r-corrected}
\end{align}

\begin{figure}[t]
    \centering
    \includegraphics[scale=1.00]{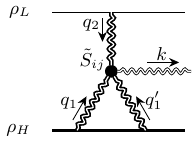}
    \caption{Diagrammatic representation of the contribution from the source term $S_{ij}^{(1,2)}$ to gravitational radiation at order $O(\mu_L\mu_H^2)$}
    \label{fig:1}
\end{figure}
It is useful to decompose the various transverse momenta into components parallel and perpendicular to the final graviton momentum. Defining
\begin{align}
\hat{\boldsymbol k}
&\equiv \frac{\boldsymbol k}{|\bsk|},
&
\hat n_i
&\equiv
\epsilon_{ji}\hat k_j,
\qquad
\epsilon_{12}=+1,
\\
q_{a\parallel}
&\equiv
\boldsymbol q_a\cdot\hat{\boldsymbol k},
&
q_{a\perp}
&\equiv
\boldsymbol q_a\cdot\hat{\boldsymbol n},
\qquad a=1,1'~,
\end{align}
and introducing the two symmetric traceless tensors
\begin{align}
\label{eq:basis-tensors}
E_{ij}
&\equiv
\hat k_i\hat k_j-\hat n_i\hat n_j
=
2\hat k_i\hat k_j-\delta_{ij}~,\qquad
F_{ij}\equiv
\hat k_i\hat n_j+\hat n_i\hat k_j\,,
\end{align}
we find that upon substituting Eqs.~\eqref{eq:alpha-corrected} and
\eqref{eq:r-corrected} into Eq.~\eqref{Sij-FT}, all inverse
powers of $\bsk$ cancel within $\widetilde S_{ij}$, and one is left with the compact expression 
\begin{align}
\widetilde S_{ij}
=
A\,E_{ij}+B\,F_{ij}\,,
\label{eq:Sij-simple}
\end{align}
where
\begin{align}
A=-\frac{\bsk^2}{2} + 2q_{1\perp}q_{1'\perp}~,~\qquad B = |\bsk|(q_{1\perp}+q_{1'\perp}) + q_{1\parallel}q_{1'\perp} + q_{1'\parallel}q_{1\perp}.
\label{eq:AB-simple}
\end{align}
We observe that $\widetilde S_{ij}^{(1,2)}$ is regular as $|\bsk|\to0$.

\subsection{Contribution from radiation field at previous order}
\label{sec:rad-previous}
\begin{figure}[ht]
    \centering
    \includegraphics[scale=0.80]{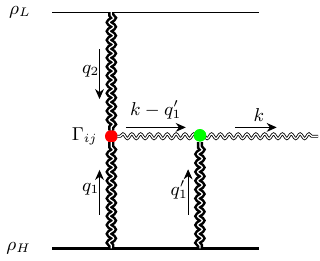}
    \caption{Rescattering correction to dilute-dense gravitational radiation in Eq.~\eqref{prev-order-FT}. The  red and green dots are, respectively, the two effective Lipatov vertices in Eq.~\eqref{eq:Gamma-sym}.}
    \label{fig:rescattering-Lipatov}
\end{figure}

The second contribution at $O(\mu_L\mu_H^2)$ arises from the
rescattering of the $O(\mu_L\mu_H)$ radiation field off the dense
shockwave,
\begin{align}
\bar g_{--}\partial_+^2\tilde h_{ij}^{(1,1)}
\big|_{\rm F.T.}(k)
={}&
2\kappa^2\mu_H(k^-)^2
\int\frac{d^2\boldsymbol q_1'}{(2\pi)^2}
\frac{dq_1^{\prime +}}{2\pi}
\frac{\tilde\rho_H(\boldsymbol q_1')}{\boldsymbol q_1'^2}
\tilde h_{ij}^{(1,1)}
\left(
\boldsymbol k-\boldsymbol q_1',
k^+-q_1^{\prime +},
k^-
\right).
\end{align}
Substituting the result obtained in Eq.~\eqref{lipatov-vertex} gives 
\begin{align}
\bar g_{--}\partial_+^2\tilde h_{ij}^{(1,1)}&
\big|_{\rm F.T.}(k)
=4\kappa^6\mu_L\mu_H^2
\int
\frac{
d^2\boldsymbol q_1'
d^2\boldsymbol q_1
d^2\boldsymbol q_2
}{(2\pi)^4}
\frac{dq_1^{\prime +}}{2\pi}
\delta^{(2)}\left(\boldsymbol k-\boldsymbol q_1'-\boldsymbol q_1-\boldsymbol q_2\right)
\nonumber\\
&\times
\frac{(k^-)^2}{
2k^-(k^+-q_1^{\prime +})
-(\boldsymbol k-\boldsymbol q_1')^2
+i\epsilon k^-
}\frac{\tilde\rho_H(\boldsymbol q_1')}{\boldsymbol q_1'^2}
\frac{\tilde\rho_H(\boldsymbol q_1)}{\boldsymbol q_1^2}
\frac{\tilde\rho_L(\boldsymbol q_2)}{\boldsymbol q_2^2}
\,
\Gamma_{ij}
\left(
\boldsymbol q_1,\boldsymbol q_2;
\boldsymbol\ell_1
\right),
\label{eq:previous-order-before-contour}
\end{align}
where $\boldsymbol\ell_1
=
\boldsymbol q_1+\boldsymbol q_2
\equiv
\boldsymbol k-\boldsymbol q_1'$, is the transverse momentum of the intermediate radiated graviton emitted from the Lipatov vertex. As illustrated in Fig.~\ref{fig:rescattering-Lipatov}, the first term in the second line above corresponds to graviton propagator in the denominator with the graviton-reggeon-graviton vertex in the numerator represented by the green blob. The other three terms in the second line correspond to the reggeon propagators. The red blob, as previously, denotes the Lipatov vertex. 

The $q_1^{\prime +}$ integral can be evaluated by closing the contour in the upper half-plane since the pole is located at $q_{1,*}^{\prime +} = k^+
-\frac{(\boldsymbol k-\boldsymbol q_1')^2}{2k^-}
+\frac{i\epsilon}{2}$. Performing the integral, we obtain 
\begin{align}
\label{prev-order-FT}
\bar g_{--}\partial_+^2\tilde h_{ij}^{(1,1)}
\big|_{\rm F.T.}(k)
&=
4\kappa^6\mu_L\mu_H^2
\left(\frac{k^-}{2i}\right)
\int
\frac{
d^2\boldsymbol q_1'
d^2\boldsymbol q_1
d^2\boldsymbol q_2
}{(2\pi)^4}
\delta^{(2)}
\left(
\boldsymbol k-\boldsymbol q_1'
-\boldsymbol q_1-\boldsymbol q_2
\right)
\nonumber\\
&\times
\frac{\tilde\rho_H(\boldsymbol q_1')}{\boldsymbol q_1'^2}
\frac{\tilde\rho_H(\boldsymbol q_1)}{\boldsymbol q_1^2}
\frac{\tilde\rho_L(\boldsymbol q_2)}{\boldsymbol q_2^2}
\Gamma_{ij}
\left(
\boldsymbol q_1,\boldsymbol q_2;
\boldsymbol\ell_1
\right).
\end{align}

Since the remaining integration measure is invariant under
$\boldsymbol q_1\leftrightarrow\boldsymbol q_1'$, we can replace $\Gamma_{ij}$ with the symmetrized kernel
\begin{align}
\Gamma^{\rm sym}_{ij}
\equiv
\frac12
\left[
\Gamma_{ij}
\left(
\boldsymbol q_1,\boldsymbol q_2;
\boldsymbol\ell_1
\right)
+
\Gamma_{ij}
\left(
\boldsymbol q_1',\boldsymbol q_2;
\boldsymbol\ell_{1'}
\right)
\right],
\label{eq:Gamma-sym}
\end{align}
where $\boldsymbol\ell_{1'}=\boldsymbol k-\boldsymbol q_1=\boldsymbol q_1'+\boldsymbol q_2~,
$  and for completeness, the gravitational Lipatov vertex entering
Eq.~\eqref{eq:Gamma-sym} is
\begin{align}
\label{eq:full-rescattering-vertex}
\Gamma_{ij}
(\boldsymbol q_a,\boldsymbol q_2;\boldsymbol\ell_a)
=&
2
\left(
q_{2i}
-\ell_{ai}\frac{\boldsymbol q_2^2}{\boldsymbol\ell_a^2}
\right)
\left(
q_{2j}
-\ell_{aj}\frac{\boldsymbol q_2^2}{\boldsymbol\ell_a^2}
\right)-
2\ell_{ai}\ell_{aj}
\frac{
\boldsymbol q_a^2\boldsymbol q_2^2
}{
\boldsymbol\ell_a^4
},
\qquad a=1,1'.
\end{align}

\subsection{Complete rescattering correction to the Lipatov vertex at
\label{sec:complete-rescattering}
$O(\mu_L\mu_H^2)$}
\begin{figure}[ht]
\centering
$$
\vcenter{\hbox{
    \includegraphics[width=0.2\textwidth]{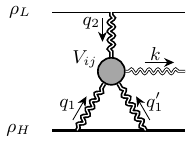}
}}
\;=\;
\vcenter{\hbox{
    \includegraphics[width=0.2\textwidth]{lipatov-rescatter-lab.pdf}
}}
\;+\;
\vcenter{\hbox{
    \includegraphics[width=0.2\textwidth]{Sij-lab.pdf}
}}
$$
\caption{Decomposition of the effective-vertex diagram for gravitational radiation at 
$O(\mu_L\mu_H^2)$ into the two terms represented in Eq.~\eqref{final-result}.}
\label{fig:effective-vertex-decomposition-3}
\end{figure}
We can now combine the source contribution
$\widetilde S_{ij}$ with the rescattering contribution
Eq.~\eqref{prev-order-FT} into a compact final result at $O(\mu_L\mu_H^2)$ that reads 
\begin{align}
\label{final-result}
\tilde h_{ij}^{(1,2)}(k)
&=
\frac{
2\mu_L\mu_H^2\kappa^6(-ik^-)
}{
k^2+i\epsilon k^-
}
\int
\frac{
d^2\boldsymbol q_1
d^2\boldsymbol q_1'
d^2\boldsymbol q_2
}{(2\pi)^4}
\delta^{(2)}
\left(
\boldsymbol k-\boldsymbol q_1
-\boldsymbol q_1'-\boldsymbol q_2
\right)
\nonumber\\
&\times
\frac{\tilde\rho_H(\boldsymbol q_1)}{\boldsymbol q_1^2}
\frac{\tilde\rho_H(\boldsymbol q_1')}{\boldsymbol q_1'^2}
\frac{\tilde\rho_L(\boldsymbol q_2)}{\boldsymbol q_2^2}
\,
V_{ij}
(\boldsymbol q_1,\boldsymbol q_1',\boldsymbol q_2)\,,
\end{align}
where the ``three reggeized graviton--graviton" vertex $V_{ij}$ in a manifestly symmetric form in $\bsq_1, \bsq_1'$ is
\begin{align}
V_{ij} = \frac{\boldsymbol q_2^2}{\bsk^2}\,
\widetilde S_{ij} - \Gamma^{\rm sym}_{ij}.
\label{eq:Vij-final}
\end{align}

In the remainder of this section, we will analyze this result further. We first look at the definite helicity projections of this vertex.  With the polarization tensor of Eq.~\eqref{polarization-tensor},
\begin{align}
\epsilon^{(-2)}_{ij}E_{ij}=
e^{-2i\phi_k}~,\qquad
\epsilon^{(-2)}_{ij}F_{ij}=-i\,e^{-2i\phi_k}\,,
\end{align}
so that its projection on Eq.~\eqref{eq:Sij-simple} becomes 
\begin{align}
\widetilde S^{(-2)}
\equiv
\epsilon^{(-2)}_{ij}\widetilde S_{ij}
=
e^{-2i\phi_k}(A-iB)\,,
\label{eq:S-helicity}
\end{align}
where $A$ and $B$ were defined previously in Eq.~\eqref{eq:AB-simple}. Likewise, the helicity projection of each Lipatov vertex was derived earlier in Eq.~\eqref{eq:temp-1}:
\begin{align}
\Gamma_a^{(-2)}
\equiv
\epsilon^{(-2)}_{ij}
\Gamma_{ij}
(\boldsymbol q_a,\boldsymbol q_2;\boldsymbol\ell_a)
=
\frac{
(\tilde q_2^*)^2\tilde q_a^{\,2}
-\boldsymbol q_a^2\boldsymbol q_2^2
}{
\tilde\ell_a^{\,2}
},
\qquad a=1,1'.
\label{eq:Gamma-helicity-rescatter}
\end{align}
The full negative-helicity vertex for Eq.~\eqref{eq:Vij-final} is therefore 
\begin{align}
V^{(-2)} = \frac{\boldsymbol q_2^2}{\bsk^2}
e^{-2i\phi_k}(A-iB) - \frac12 \left( \Gamma_1^{(-2)} + \Gamma_{1'}^{(-2)}
\right).
\label{eq:V-helicity-final}
\end{align}
Taking the residue of the outgoing graviton pole, the 
$O(\mu_L\mu_H^2)$ negative helicity Lipatov radiation amplitude is 
\begin{align}
\mathcal M_{\rm Lipatov}^{(1,2),(-2)}(k)
\equiv
\lim_{k^2\to0}
\left(
k^2+i\epsilon k^-
\right)
\frac{1}{\kappa}\epsilon^{(-2)}_{ij}
\tilde h_{ij}^{(1,2)}(k)\,,
\end{align}
or equivalently, 
\begin{align}
\mathcal M_{\rm Lipatov}^{(1,2),(-2)}(k)
=
2\kappa^5\mu_L\mu_H^2(-ik^-)\!\!
\int
\frac{
d^2\boldsymbol q_1
d^2\boldsymbol q_1'
d^2\boldsymbol q_2
}{(2\pi)^4}
\delta^{(2)}
\left(
\boldsymbol k-\boldsymbol q_1
-\boldsymbol q_1'-\boldsymbol q_2
\right)
\frac{
\tilde\rho_H(\boldsymbol q_1)
\tilde\rho_H(\boldsymbol q_1')
\tilde\rho_L(\boldsymbol q_2)
}{
\boldsymbol q_1^2
\boldsymbol q_1'^2
\boldsymbol q_2^2
}
V^{(-2)}.
\label{eq:M-Regge-12}
\end{align}

For the positive helicity amplitude\footnote{
The positive helicity polarization tensors are
\begin{align}
\epsilon_{\mu}^{(+)}
=
\frac{1}{\sqrt{2}}(0,1,+i,0),
\qquad
\epsilon_{\mu\nu}^{(+2)}
=
\epsilon_{\mu}^{(+)}\epsilon_{\nu}^{(+)} ~.
\end{align}
}, performing similarly the contractions with the basis tensors $E_{ij}$ and $F_{ij}$, the positive helicity $\widetilde S^{(+2)}$ contribution is 
\begin{align}
\widetilde S^{(+2)}
\equiv
\epsilon_{ij}^{(+2)}\widetilde S_{ij}
=
e^{+2i\phi_k}(A+iB)\,,
\end{align}
and for each Lipatov vertex, 
\begin{align}
\Gamma_a^{(+2)}
=
\frac{
(\tilde q_2)^2(\tilde q_a^*)^2
-\boldsymbol q_a^2\boldsymbol q_2^2
}{
(\tilde\ell_a^*)^2
}\,,
\qquad a=1,1'.
\end{align}
The positive-helicity effective vertex is therefore
\begin{align}
V^{(+2)}
=
\frac{\boldsymbol q_2^2}{\bsk^2}
e^{+2i\phi_k}(A+iB)
-
\frac12
\left(
\Gamma_1^{(+2)}
+
\Gamma_{1'}^{(+2)}
\right)~,
\end{align}
and when plugged into Eq.~\eqref{eq:M-Regge-12}, gives the positive helicity 
contribution $\mathcal M_{\rm Lipatov}^{(1,2),(+2)}(k)$ to the $O(\mu_L\mu_H^2)$ gravitational radiation amplitude. Since the transverse momenta are real, we have 
$V^{(+2)}=\left[V^{(-2)}\right]^*$. 

From Eq.~\eqref{eq:M-Regge-12}, we can extract the contribution to the large frequency $\omega$ behaviour of the spectrum $dE_{\text{GW}}/d\omega$ at this order. First note from Eqs.~\eqref{eq:Vij-final}, \eqref{eq:full-rescattering-vertex} and \eqref{eq:Sij-simple} that at large $\omega$ the dominant contribution comes from the $\tilde{S}_{ij}$ part of the full vertex $V_{ij}$. This is because at large $\omega$ the rescattering term from the previous order is suppressed as $1/\omega^2$ whereas the source term is constant in $\omega$. 
In the impact parameter representation, the momentum integrals involving the source contribution to the amplitude can be evaluated in closed form. We can write  Eq.~\eqref{eq:M-Regge-12} as
\begin{align}
\mathcal{M}^{(1,2),(-2)}_{\text{Lipatov}}=2 \kappa^5 \mu_L \mu_H^2 \left(-i k^{-}\right)\left[\mathcal{I}_{\rm source}+\mathcal{I}_{\text {rescat. }}\right] .
\end{align}
We established previously that $\mathcal{I}_{\text {rescat. }}\sim 1/\omega^2$ at large $\omega$, so we will forcus on the source term. In  the point particle limit, with $\tilde\rho_H(\boldsymbol q_1)
=\tilde\rho_H(\boldsymbol q_1')=1$, and  $\tilde\rho_L(\boldsymbol q_2)
=e^{-i\boldsymbol q_2\cdot\boldsymbol b}$,  
\begin{align}
\mathcal{I}_{\rm source} &= \int
\frac{
d^2\boldsymbol q_1
d^2\boldsymbol q_1'
d^2\boldsymbol q_2
}{(2\pi)^4}
\delta^{(2)}
\left(
\boldsymbol k-\boldsymbol q_1
-\boldsymbol q_1'-\boldsymbol q_2
\right)
\frac{
\tilde\rho_H(\boldsymbol q_1)
\tilde\rho_H(\boldsymbol q_1')
\tilde\rho_L(\boldsymbol q_2)
}{
\boldsymbol q_1^2
\boldsymbol q_1'^2
\boldsymbol q_2^2
}\frac{\boldsymbol q_2^2}{\bsk^2}
e^{-2i\phi_k}(A-iB)~,\nonumber\\[5pt]
&=e^{-i \bsk \cdot \bsb} e^{-2 i \phi_k}\left[\frac{e^{2 i \Delta \phi}-1}{4 \pi^2 |\bsk|^2 b^2}-\frac{\log (\mu b)}{2 \pi^2 |\bsk|} \frac{\sin \Delta\phi}{b}-\frac{\log ^2(\mu b)}{8 \pi^2}\right] .
\end{align}
Here $\mu$ in an IR cutoff, $\Delta\phi=\phi_b-\phi_k$ and $b$ is the modulus of the impact parameter $\bsb$. We observe that for large $\omega$,  
\be
\mathcal{I}_{\rm source}=O(1)+O\left(\omega^{-1}\right)+O\left(\omega^{-2}\right)\,.
\ee
and therefore multiplying by an explicit $k^{-}$ factor ($\sim \omega \theta^2$ for fixed small angle) has the structure 
$$
\mathcal{M}_{\text{Lipatov}}^{(1,2)}=O(\omega)+O(1)+O\left(\omega^{-1}\right)~,
$$
with the leading term in the large $\omega$ limit being 
\begin{align}
\mathcal{M}_{\text{Lipatov}}^{(1,2),(-2)} \simeq \frac{i \kappa^5\mu_L \mu_H^2 }{8 \sqrt{2} \pi^2} \omega \theta^2 \log ^2(\mu b) e^{-i \omega b \theta \cos \Delta \phi} e^{-2 i \phi_k}~.
\end{align}
Clearly in the large $\omega$ regime, the leading contribution from the $O(\mu_L\mu_H^2)$ term is greater than the $O(\mu_L\mu_H)$ contribution discussed above Eq.~\eqref{eq:LO-large-omega}. This implies that an all-order resummation in the dilute-dense framework is needed before we can take the large frequency limit to determine the UV behavior of the spectrum. The soft limit of $\omega\sim0$ on the other hand does not suffer from this issue, as we will now discuss.

\subsection{Soft limit of Lipatov amplitude at $O(\mu_L\mu_H^2)$}
\label{sec:Lipatov-soft}
We next study the limit $\boldsymbol{k}^2\to0$ for fixed direction $\hat{\boldsymbol k}$. From Eqs.~\eqref{eq:AB-simple}, we have
\begin{align}
A=2q_{1\perp}q_{1'\perp}
+\mathcal O(\bsk^2)~,\qquad
B=q_{1\parallel}q_{1'\perp}+q_{1'\parallel}q_{1\perp}+\mathcal O(|\bsk|)~.
\end{align}
Therefore
\begin{align}
\widetilde S^{(-2)}
=
e^{-2i\phi_k}
\Big[
2q_{1\perp}q_{1'\perp}
-i\big(
q_{1\parallel}q_{1'\perp}
+
q_{1'\parallel}q_{1\perp}
\big)
\Big]
+\mathcal O(|\bsk|)~.
\label{eq:S-soft}
\end{align}
For generic finite transverse momentum transfers $\bsq_1$ and $\bsq_1'$, $\Gamma^{(-2)}_{1}$ and $\Gamma^{(-2)}_{1'}$ remain finite in this limit; this follows from the definition of $\ell_a$. Hence, 
\begin{align}
V^{(-2)}_{\text{soft}} =
\frac{\boldsymbol q_2^2}{\bsk^2}
e^{-2i\phi_k}
\Big[
2q_{1\perp}q_{1'\perp}
-i\big(
q_{1\parallel}q_{1'\perp}
+
q_{1'\parallel}q_{1\perp}
\big)
\Big]
+\mathcal O(|\bsk|^{-1})~.
\label{eq:V-soft}
\end{align}
The factor $\boldsymbol q_2^2$ cancels the light-source propagator in
Eq.~\eqref{eq:M-Regge-12}; the leading soft pole is generated entirely by the corrected source term.

We next evaluate the leading soft term explicitly for pointlike
transverse sources, with $\tilde\rho_H(\boldsymbol q_1)
=\tilde\rho_H(\boldsymbol q_1')=1$, and  $\tilde\rho_L(\boldsymbol q_2)
=e^{-i\boldsymbol q_2\cdot\boldsymbol b}$. We then get 
\begin{align}
\mathcal M_{\rm soft}^{(1,2),(-2)}
=
\frac{
2\kappa^5\mu_L\mu_H^2(-ik^-)
e^{-2i\phi_k}}{\bsk^2}
\int
\frac{
d^2\boldsymbol q_1
d^2\boldsymbol q_1'
d^2\boldsymbol q_2
}{(2\pi)^4}
\delta^{(2)}
\left(
\boldsymbol k-\boldsymbol q_1
-\boldsymbol q_1'-\boldsymbol q_2
\right)
e^{-i\boldsymbol q_2\cdot\boldsymbol b}\,\,
\frac{
{\cal N}(\boldsymbol q_1,\boldsymbol q_1')
}{
\boldsymbol q_1^2\boldsymbol q_1'^2
}~,
\label{eq:M-soft-before-integrals}
\end{align}
where ${\cal N}\equiv
2q_{1\perp}q_{1'\perp}
-i\left(
q_{1\parallel}q_{1'\perp}
+
q_{1'\parallel}q_{1\perp}
\right)$. 
Performing the $\boldsymbol q_2$ integral, and using 
$e^{-i\boldsymbol k\cdot\boldsymbol b} =1+\mathcal O(|\bsk|)$, we then get 
\begin{align}
\mathcal M_{\rm soft}^{(1,2),(-2)}
={}&
\frac{
2\mu_L\mu_H^2\kappa^5(-ik^-)
}{\bsk^2}
e^{-2i\phi_k}
\int
\frac{d^2\boldsymbol q_1}{(2\pi)^2}
\frac{e^{i\boldsymbol q_1\cdot\boldsymbol b}}
{\boldsymbol q_1^2}
\int
\frac{d^2\boldsymbol q_1'}{(2\pi)^2}
\frac{e^{i\boldsymbol q_1'\cdot\boldsymbol b}}
{\boldsymbol q_1'^2}
\,{\cal N}~.
\label{eq:M-soft-factorized}
\end{align}
The two remaining transverse momentum integrations give
\begin{align}
\int\frac{d^2\boldsymbol q_1}{(2\pi)^2}
\frac{e^{i\boldsymbol q_1\cdot\boldsymbol b}}
{\boldsymbol q_1^2}
\int
\frac{d^2\boldsymbol q_1'}{(2\pi)^2}
\frac{e^{i\boldsymbol q_1'\cdot\boldsymbol b}}
{\boldsymbol q_1'^2}
\,{\cal N}=
-\frac{b_\perp^2}{2\pi^2b^4}
+
i\frac{b_\parallel b_\perp}{2\pi^2b^4}\equiv \frac{1}{4\pi^2b^2}
\left(
e^{2i\Delta\phi}-1
\right)\,,
\label{eq:soft-double-FT}
\end{align}
where $b_\parallel=b\cos\Delta\phi$, $b_\perp=b\sin\Delta\phi$, and 
$\Delta\phi=\phi_b-\phi_k$. 

Substituting this result into
Eq.~\eqref{eq:M-soft-factorized}, we obtain the soft limit of the Lipatov amplitude to 
$O(\mu_L\mu_H^2)$ to be\footnote{We used here $2k^+k^-=\bsk^2$ and 
$k^+ = \frac{\omega}{\sqrt2}\left(1+\sqrt{1-\theta^2}\right) = \sqrt2\,\omega
+\mathcal O(\omega\theta^2)$.}
\begin{align}
\mathcal M_{\rm soft}^{(1,2),(-2)}
(\boldsymbol b;k)
=-\frac{
i\kappa^5\mu_L\mu_H^2
}{4\sqrt{2}\pi^2 b^2}
\frac{1}{\omega}\,
e^{-2i\phi_k}
\left(
e^{2i\Delta\phi}-1
\right)\,.
\label{eq:M-soft-b-omega}
\end{align}
This result for the soft limit of the $O(\mu_L\mu_H^2)$  Lipatov amplitude recovers the expected leading Weinberg behavior $\mathcal M_{\rm soft}^{(1,2)}
\sim \frac{1}{\omega}$, 
which should hold true for all orders in the dilute-dense expansion. This result also shows that the soft limit of this Regge amplitude vanishes identically when the graviton emission is in the direction of the impact parameter $(\Delta \phi=0)$  which, as we showed previously, was also the case for the $O(\mu_L\mu_H)$ result. As an independent check of this result, we derive in Appendix~\ref{app:eikonal-soft-NLO} derive Eq.~\eqref{eq:M-soft-b-omega} from the connected part of the two-graviton exchange eikonal Feynman diagrams in the Weinberg soft limit.

\subsection{Comment on the dilute-dense QCD-GR double copy} 
\label{sec:comments}
Our result can be compared to the corresponding result for the QCD case we noted earlier in Eq.~\eqref{eq:dilute-denseQCD}:
\begin{align}
&a_i(k)=-\frac{2 i g}{k^2+i \epsilon k^{-}} \int \frac{d^2 \boldsymbol{q}_2}{(2 \pi)^2}\left(q_{2 i}-k_i \frac{\boldsymbol{q}_2^2}{\boldsymbol{k}^2}\right)\frac{\tilde{\rho}_L\left(\boldsymbol{q}_2\right)}{\boldsymbol{q}_2^2}\left(U\left(\boldsymbol{k}+\boldsymbol{q}_2\right)-(2 \pi)^2 \delta^2\left(\boldsymbol{k}+\boldsymbol{q}_2\right)\right)~.\nonumber
\end{align}
Recall that here the expression in the brackets $\left(q_{2 i}-k_i \frac{\boldsymbol{q}_2^2}{\boldsymbol{k}^2}\right)$ is the QCD Lipatov vertex, and  $U(\boldsymbol{k})$  is the Fourier transform of the lightlike Wilson line operator
\begin{align}
\label{gauge-WL}
U\left(x^{-}, \boldsymbol{x}\right) \delta\left(x^{+}\right)=\exp \left(i g \int_{-\infty}^{x^{-}} d z^{-} \bar{A}_{-}\left(z^{-}, \boldsymbol{x}\right) \cdot T\right)\,,\nonumber
\end{align}
where $\bar{A}_\mu\left(x^{-}, \boldsymbol{x}\right)=-g \delta_{\mu-} \delta\left(x^{-}\right) \frac{\rho_H(\boldsymbol{x})}{\square_{\perp}}$.

A priori, it is not obvious if a similar form exists in GR, where an all order resummation in $\mu_H$ leads to a factorization between the gravitational Lipatov vertex and the gravitational Wilson line. The latter implicitly appears in the formal exact solution that we wrote in Section \ref{sec:exact-solution} via the retarded shockwave propagator in Eq.~\eqref{T-tensor}. The exponential in that equation can be written analogously to the above expression as 
\be
\label{grav-WL}
U_k(\boldsymbol{x})=\exp \left(i \kappa^2 \mu_H \frac{\rho_H(\boldsymbol{x})}{\square_{\perp}} k^{-}\right)=P \exp \left(\frac{i}{2} \int d z^{-} \bar{g}_{--}\left(z^{-}, \boldsymbol{x}\right) k^{-}\right)\,.
\ee
Notice that if we were to expand the exponential above, it will bring down additional factors of $ik^-$. We see this in our result Eq.~\eqref{final-result}, where a $ik^-$ sits upfront. Note further that the this factor is common to both $\tilde{S}_{ij}$ and $\Gamma_{ij}$ in Eq.~\eqref{final-result}. It appears in $\tilde{S}_{ij}$ terms manifestly in the Fourier transforms (see Eqs. \eqref{line1}, \eqref{line2}, \eqref{line3} where an additional factor of $ik^+$ hangs in the denominator that we later replaced with $ik^-/\bsk$ via the on-shell condition on the 4-momenta of the graviton $k$. On the other hand, a factor of $ik^-$ emerges  in the $\Gamma_{ij}$ term from performing the $q_1'^{+}$ integral over the intermediate graviton propagator. While this observation hints, as in the QCD case, at a possible resummation of all order contributions in $\mu_H$ into the gravitational Wilson line, this is not guaranteed, and is an important test of the double copy

\section{Summary and outlook}
\label{sec:Conclusions}
In this work, we formulated the problem of gravitational radiation in trans-Planckian shockwave scattering of a light mass compact object with energy $\mu_L$  off a heavy compact object with energy $\mu_H$ in a dilute-dense framework where $\mu_L\ll \mu_H$. Our framework is exactly analogous to the computation of gluon radiation from Yang-Mills shockwaves. We showed that the solution to all orders can be expressed in terms of shockwave propagators that are a double copy of the Yang-Mills propagators. The graviton shockwave propagators are all-order expressions in $\mu_H$  or equivalently, in the transverse mass density $\rho_H$ of the heavy object and the gravitational coupling. The radiative energy spectrum for semi-hard graviton frequencies $0\ll \omega \ll \sqrt{s}$ (with $s=2\mu_L\mu_H$) is therefore sensitive to multipoint correlators of $\rho_H$ that encode the dynamical response of the heavy object. These provide fundamental information on classical, and potentially semi-classical, effects in black hole formation as $b\rightarrow R_S$~\ \cite{Raj:2025hse}. As discussed previously~\cite{Stasto:2026xqw}, the latter may be enhanced significantly if the radiation is in the form of a Susskind-Glogower-type squeezed state. 

Due to the complexity of the expressions, we are unable at present to extract the all-order radiation spectrum directly from the our formal solution. We were instead able to formulate the problem as a hierarchy of gravitational field equations that can be solved order-by-order in the dilute-dense expansion. To leading nontrivial order $O(\mu_L\mu_H)$, we recovered our previous result where the radiation spectrum is governed by the gravitational Lipatov vertex. This vertex is proportional to the double copy of the QCD Lipatov vertex, which is a key building block of scattering in the Regge asymptotics of QCD. We showed that the additional double copy dependence of the gravitational Lipatov vertex on the QED bremsstrahlung vertex is required by a careful smearing of the transverse mass densities when going beyond the point particle approximation. In the scattering amplitude language, this term was shown by Lipatov to be required by unitarity. 

We examined further the angular structure of Lipatov radiation. This enabled us to recover straightforwardly our previous result that the soft limit of the Lipatov amplitude gives the ultrarelativistic limit of the Weinberg amplitude for single graviton exchange. In the Weinberg $\omega\rightarrow 0$ limit, the frequency dependence of the spectrum goes to a constant; in contrast, we see that in the semi-hard region the spectrum behaves as $1/\omega^2$. 

We next solved the dilute-dense hierarchy explicitly to order $O(\mu_L\mu_H^2)$. The result can be expressed as a sum of two terms. The first corresponds precisely to the rescattering of the graviton emitted from the Lipatov vertex off a reggeized gluon from the dense source while the second corresponds to graviton emission from a vertex arising from the fusion of three reggeized contributions. In the large $\omega$ limit, the latter contribution is dominant, with the leading term in the amplitude being proportional to $\omega$. However since this is a fixed order computation, it is premature to ignore the rescattering contribution and one needs to work to at least one higher order to get a clearer sense of how the hierarchy may resum. Though somewhat cumbersome, the $O(\mu_L\mu_H^3)$ result can be determined and will be reported on in follow-up work. A further important element will be to understand the underlying structure of the response functions and how they can impact the radiation spectrum. 

In Appendix \ref{app:CCV}, we examined the eikonal multiple scattering framework of Veneziano and collaborators. We see in that case the Weinberg limit does not emerge as a smooth limit of the Lipatov amplitude but that there is an intermediate frequency regime where the two results need to be matched explicitly. In the shockwave case, the Weinberg soft result for ultrarelativistic scattering is a limit of the Lipatov expression. As we show explicitly in Appendix \ref{app:eikonal-soft-NLO}, this holds even at $O(\mu_L\mu_H^2)$, where the soft limit of our result corresponds to radiation from the incoming and outgoing light particle with two graviton exchanges with the heavy particle sandwiched in between.  The origin of this discrepancy with CCCV needs to be better understood and will also be addressed in future work. 

Our study of their work in Appendix \ref{app:CCV}  was especially helpful because it suggests that it is important in general  to go beyond the strict shockwave limit of $\gamma\rightarrow \infty$ and systematically include $1/\gamma$ corrections that can modify the gravitational wave spectrum. At finite $\gamma$, there is for instance an interesting interplay between the coherence time of the scattering and the mean free path between collisions that becomes important. This is the gravitational analog of the Landau-Pomeranchuk-Migdal (LPM) effect in QED~\cite{Landau:1965ksp,Migdal:1956tc}, and in QCD~\cite{Baier:2000mf}, 
that causes a destructive interference, potentially qualitatively altering  the frequency spectrum in a wide kinematic range. The gravitational LPM effect will also be addressed in follow-up work.  

\section*{Acknowledgments}
We would like to thank Tim Adamo, Ratindranath Akhoury, Leor Barack, Fabian Bautista, Gia Dvali, Mathieu Giroux, Riccardo Gonzo, Anton Ilderton, Donal O'Connell, Adam Pound, Radu Roiban, Ira Rothstein, Anna Stasto and Gabriele Veneziano for useful comments that have influenced this work.  I. B. (n\'{e}e Maria Isabelle Fite) was supported in part by the Brookhaven National Laboratory (BNL) Physics Department under the BNL Supplemental Undergraduate Research Program (SURP). I. B.  would also like to acknowledge support from the National Science Foundation through the Oklahoma Louis-Stokes Alliance for Minority Participation Program. H. R. is a Simons Foundation Post-doctoral Fellow at Stony Brook supported under Award number 994318. R. V. is supported at Stony Brook by the Simons Foundation as a co-PI under Award number 994318 (Simons Collaboration on Confinement and QCD Strings). R. V. is supported at BNL  by the U.S. Department of Energy, Office of Science under contract DE-SC0012704 and within the framework of the Saturated Glue (SURGE) Topical Collaboration in Nuclear Theory.  R. V. thanks the UK Royal Society and the Wolfson Foundation for a Visiting Fellowship and the Higgs Center at the Univ. of Edinburgh for their kind hospitality. 

\appendix

\section{Matching Weinberg to Lipatov in the CCCV formalism}
\label{app:CCV}

To extract the $\omega$-dependence of the spectrum , we will {\it a la} CCCV, we will employ their \emph{soft-Regge matching}, which makes transparent how soft Weinberg radiation is embedded within the Lipatov radiation spectrum. The soft amplitude we will construct from Weinberg currents and the Regge amplitude in Eq.~\eqref{app-M1-b} are valid in different but overlapping
kinematic regions of the $(\bth,\bth_s)$ plane:
\begin{itemize}\setlength\itemsep{2pt}
\item Region (a): $|\bth_s|\gg|\bth|$ ---
\emph{only soft is valid}; the emission is from external lines.
\item Region (b): $|\bth|\gg|\bth_s|\gg\tfrac{\omega}{E}|\bth|$ ---
\emph{both soft and Regge are valid and equivalent}, since the
sub--energies are entering Regge kinematics but external line
insertions still dominate.
\item Region (c): $|\bth_s|\ll \tfrac{\omega}{E}|\bth|$ ---
\emph{only Regge is valid}; the soft approximation breaks down and the radiation includes emission from the exchanged graviton line as well.
\end{itemize}
\noindent
In region (b), where both representations are valid, they must give the
same answer. CCCV verify this directly by computing the
difference $\mathcal{M}_{\text{Regge}}-\mathcal{M}_{\text{soft}}$ and showing it
vanishes in the soft regime (see also \cite{Lipatov:1982it, Raj:2025hse}). The \emph{soft-Regge matching} exploits this overlap to glue the two representations into a single unified amplitude which is valid in all of (a)$\cup$(b)$\cup$(c).

More specifically, the matching works as follows. We consider the soft amplitude as a starting point
\begin{align}
\label{eq:Msoft-base}
\mathcal{M}_{\text{soft}}(b;E,\omega,\bth)
\;=\; s\kappa^3\,\frac{E}{\omega}
\int\!\frac{d^2\bth_s}{(2\pi)^2|\bth_s|^2}\,
e^{iE\bsb\cdot\bth_s}\,e^{-2i\fp_\theta}\,
\bigl(e^{2i\fp_{\theta-\theta_s}}-e^{2i\fp_\theta}\bigr).
\end{align}
The above equation is derived in a manner that is similar to the derivation of the Regge amplitude in Eq.~\eqref{app-M1-b} where we replace the Lipatov vertex with the Weinberg vertex. Writing this amplitude as
\begin{equation}
\mathcal{M}_{\text{soft}}(b;E,\omega,\bth) \;=\; s\kappa^3 \int\!\frac{d^2\qv}{(2\pi)^2}\;
e^{i\qv\cdot\bv}\;\mathcal{I}_{\text{soft}}(\kv,\qv) \ ,
\label{mtc:normalisation-1}
\end{equation}
where $\qv = E\ths$ and $\kv \;=\; \omega\tht$, with the integrand $\mathcal{I}_{\text{soft}}$
\begin{align}
\mathcal{I}_{\text{soft}}(\kv,\qv)  = \frac{E}{\omega}\frac{1}{|\bsq_2^2|} \left(e^{2i(\varphi_{\theta-\theta_s}-\varphi_\theta)} - 1 \right)
&\;=\; \frac{E}{\omega}\frac{1}{|\bsq_2^2|} \left(\frac{(\tilde \theta-\tilde \theta_s)\,\tilde \theta^{*}}{(\tilde \theta-\tilde \theta_s)^{*}\,\tilde \theta} - 1 \right)\ ,\nonumber \\[5pt]
&\;=\; \frac{E}{\omega}\frac{1}{|\bsq_2^2|} \left( \frac{\tilde \theta\tilde \theta_s^{*} - \tilde \theta^{*}\tilde \theta_s}{\tilde \theta\,(\tilde \theta-\tilde \theta_s)^{*}}\right)~.
\label{mtc:softbracket1}
\end{align}
In terms of complex $2d$ momenta, the numerator and denominator of the above expression on the right is
\begin{align}
\tilde \theta\tilde \theta_s^{*} - \tilde \theta^{*}\tilde \theta_s &= \frac{\tilde k\tilde q_2^{*}-\tilde k^{*}\tilde q_2}{\omega E} , \qquad \tilde \theta\,(\tilde \theta-\tilde \theta_s)^{*} = \frac{\tilde k\,(\tilde k^{*}-x\,\tilde q_2^{*})}{\omega^2} ~.
\end{align}
Plugging this above, we get the exact expression for $\mathcal{I}_{\text{soft}}$
\begin{align}
\label{eq:I-soft}
\mathcal{I}_{\text{soft}} (\kv,\qv)  =\frac{1}{\qv^2}\; \frac{\tilde k \tilde q_2^{*}-\tilde k^{*}\tilde q_2}{\tilde k\,\big(\tilde k^{*}-x\,\tilde q_2^{*}\big)} \;=\; \frac{1}{\omega E}\frac{1}{|\bth_s^2|} \left( \frac{\tilde \theta\tilde \theta_s^{*} - \tilde \theta^{*}\tilde \theta_s}{\tilde \theta\,(\tilde \theta-\tilde \theta_s)^{*}}\right)~.
\end{align}
where we used the short hand notation $x=\omega/E$. In this form, it is trivial to see that the high energy limit of $\Isoft$ (where we take $E\to \infty$ or $x\to 0$) agrees with what was derived earlier in  Eq.~\eqref{eq:I_soft-HE} upto an overall helicity factor of $e^{-2i\varphi_k}$ which can be absorbed into the definition of the graviton polarization tensor. Further, observe that the entire difference between $\mathcal{I}_{\text{soft}}$ above and $\mathcal{I}_{\text{Regge}}$ in Eq.~\eqref{eq:IRegge} is in the denominator (the numerators are identical). To identify the regime, where these two amplitudes agree look at the ratio $\IRegge / \Isoft$ (ignoring the helicity factor of $e^{-2i\varphi_k}$):
\begin{equation}
\frac{\IRegge}{\Isoft}
= \frac{|\tilde \theta_s|^2\ \tilde \theta\,(\tilde \theta-\tilde \theta_s)^{*}}
       {|\tilde \theta|^2\ \tilde \theta_s\,(\tilde \theta_s - x\tilde \theta)^{*}}
= \frac{\tilde \theta_s^{*}(\tilde \theta-\tilde \theta_s)^{*}}{\tilde \theta^{*}(\tilde \theta_s - x\tilde \theta)^{*}}
= \left[ \frac{\tilde \theta_s\,(\tilde \theta-\tilde \theta_s)}{\tilde \theta\,(\tilde \theta_s - x\tilde \theta)} \right]^{*} .
\label{ccvm:ratio}
\end{equation}
The two amplitudes agree when the above ratio tends to 1. This implies that the angles $\tilde \theta$ and $\tilde \theta_s$ be ordered as
\begin{align}
|\bth_s| \ll |\bth|
\quad\text{and}\quad
x |\bth| \ll |\bth_s| \ ,
\end{align}
or combining them: $|\bth| \;\gg\; |\bth_s| \;\gg\; \frac{\omega}{E}\,|\bth|$ gives the bounds on region (b) defined above.

\begin{figure}[h]
  \centering
  \includegraphics[scale=1.25]{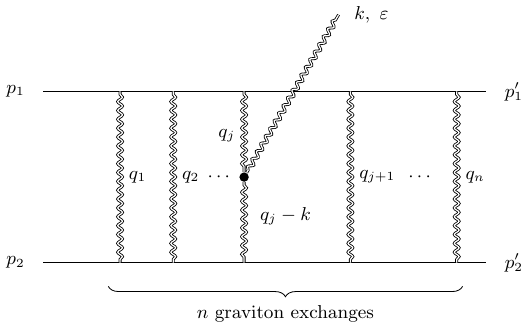}
  \caption{Single graviton emission off the eikonal ladder.}
  \label{fig:eikonal-emission}
\end{figure}

The soft amplitude Eq.~\eqref{eq:Msoft-base} is valid in region (a) and agrees with the Regge amplitude in Eq.~\eqref{app-M1-b} only in region (b). In particular its validity does not hold in region (c). To extend its validity into region (c), CCCV add to it the difference $\Delta\mathcal{M} \equiv \mathcal{M}_{\text{Regge}}-\mathcal{M}_{\text{soft}}$. After careful analysis of this
difference they find that $\Delta\mathcal{M}$ in the overlapping regime is formally equal to the negative of the soft amplitude evaluated with $E$ replaced by $\omega$. Adding this difference to the soft amplitude $\mathcal{M}_{\text{soft}}$ gives the matched amplitude:
\begin{equation}\label{eq:Mmatched-def}
\mathcal{M}_{\text{matched}} \;\equiv\; \mathcal{M}_{\text{soft}}(b;E,\omega,\bth)
\;-\;\mathcal{M}_{\text{soft}}(b;\omega,\omega,\bth)\;.
\end{equation}
Writing out Eq.~\eqref{eq:Mmatched-def} explicitly using Eq.~\eqref{eq:Msoft-base} we get
\begin{align}
\mathcal{M}_{\text{matched}}(b;E,\omega,\bth)
&= s\kappa^3 
\int\!\frac{d^2\bth_s}{(2\pi)^2|\bth_s|^2}\,
\left(\frac{E}{\omega}\,e^{iE\bsb\cdot\bth_s}
\;-\;e^{i\omega\bsb\cdot\bth_s}\right)\,
\bigl(e^{2i\fp_{\theta-\theta_s}}-e^{2i\fp_\theta}\bigr).
\label{eq:Mmatched-step1}
\end{align}
To proceed, we convert the helicity phase difference factor
$(e^{2i\fp_{\theta-\theta_s}}-e^{2i\fp_\theta)}$ into an integral over an auxiliary complex variable $\bz$ using the identity:
\begin{equation}\label{eq:A25}
e^{2i\fp_{\theta_A}} - e^{2i\fp_{\theta_B}}
\;=\;\int\!\frac{d^2 \bz}{\pi\,\bz^{*2}}\,
\bigl(e^{iA\bz\cdot\bth_B}-e^{iA\bz\cdot\bth_A}\bigr),
\end{equation}
where the variable $A$ is arbitrary. Apply Eq.~\eqref{eq:A25} with $\bth_A=\bth-\bth_s$, $\bth_B=\bth$
\begin{align}
e^{2i\fp_{\theta-\theta_s}}-e^{2i\fp_\theta}
&= \int\!\frac{d^2 \bz}{\pi\,\bz^{*2}}\,
\bigl(e^{iA\bz\cdot\bth}-e^{iA\bz\cdot(\bth-\bth_s)}\bigr)
\label{eq:phase-zrep}
\end{align}
Next, substitute Eq.~\eqref{eq:phase-zrep} into Eq.~\eqref{eq:Mmatched-step1} and interchange the order of the $\bth_s$ and $\bz$ integrations:
\begin{equation}\label{eq:Mmatched-step2}
\mathcal{M}_{\text{matched}}(b;E,\omega,\bth) = s\kappa^3
\int\!\frac{d^2 \bz}{\pi \bz^{*2}}\,e^{iA\bz\cdot\theta}\,\bigl[I_1(\bz) - I_2(\bz)\bigr],
\end{equation}
where the two $\bth_s$ integrals are
\begin{align}
I_1(\bz) &\equiv \frac{E}{\omega}\!\int\!\frac{d^2\bth_s}{(2\pi)^2|\bth_s|^2}
\,(e^{iE \bsb\cdot\bth_s}-e^{i(E \bsb-A\bz)\cdot\bth_s}),\label{eq:I1-def}\\[3pt]
I_2(\bz) &\equiv \!\int\!\frac{d^2\bth_s}{(2\pi)^2|\bth_s|^2}
\,(e^{i\omega\bsb\cdot\bth_s}-e^{i(\omega\bsb-A\bz)\cdot\bth_s}).\label{eq:I2-def}
\end{align}
Both $I_1$ and $I_2$ involve the basic integral
\begin{equation}\label{eq:basic-Fourier}
\mathcal{I}(\bsa_1, \bsa_2)\;\equiv\;
\int\!\frac{d^2\bth_s}{2\pi|\bth_s|^2}\,
\bigl(e^{i\bsa_1\cdot\bth_s}-e^{i\bsa_2\cdot\bth_s}\bigr).
\end{equation}
Each term above on its own is logarithmically IR-divergent as
$|\bth_s|\to 0$. However, their difference is finite. As a result we get
\begin{equation}\label{eq:basic-Fourier-diff}
\mathcal{I}(\bsa_1, \bsa_2) \;=\;
\log\!\frac{|\bsa_2|}{|\bsa_1|}
\end{equation}
Now apply this to $I_1$ and $I_2$ integrals. For $I_1$ we have $\bsa_1 = E\bsb $ and $\bsa_2 = E\bsb  - A\bz$:
\begin{equation}\label{eq:I1-result}
I_1(\bz) = \frac{E}{2\pi\omega}\,\log\!\frac{|E\bsb  - A\bz|}{|E\bsb |}
= \frac{E}{2\pi\omega}\,\log\!\left|\hat{\bb}-\frac{A}{E |\bsb|}\,\bz\right|~.
\end{equation}
For $I_2$, $\bsa_1 = \omega\bsb$ and $\bsa_2 = \omega\bsb - A\bz$:
\begin{equation}\label{eq:I2-result}
I_2(z) = \frac{1}{2\pi}\log\frac{|\omega\bsb-A\bz|}{|\omega b|}
= \frac{1}{2\pi}\log\left|\hat{\bb}-\frac{A}{\omega b}\,\bz\right|~.
\end{equation}
Here we used $\bsb = b\,\hat{\bb}$ so $|\bsb|=b$. We now exploit the freedom to choose $A$. A simple choice is to set
$A=\omega b$, which makes the second log argument simply
$|\hat{\bb}-\bz|$. The final result of the two integrals $I_1(\bz)$ and $I_2(\bz)$ are: 
\begin{align}
\label{eq:I-final}
I_1(\bz) &= \frac{1}{2\pi}\frac{E}{\omega}\,\log \left|\hat{\bb}-\frac{\omega}{E}\, \bz\right|~,\quad I_2(\bz) = \frac{1}{2\pi}\log \bigl|\hat{\bb}-\bz\bigr|~.
\end{align}
Next, we substitute the above result into
Eq.~\eqref{eq:Mmatched-step2}. This gives the following result of the matched amplitude. 
\begin{equation}\label{eq:M-matched-Phi-R}
\mathcal{M}_{\text{matched}}(b;E,\omega,\bth) \;=\; s\kappa^3
\int \frac{d^2 \bz}{2\pi^2 \bz^{*2}}\,e^{ib\omega z\cdot\theta}\,
\left(\,\frac{E}{\omega}\log|\hat{\bb}-(\omega/E)\bz|
\;-\;\log|\hat{\bb}-\bz|\,\right)\,.
\end{equation}
Define
\begin{equation}\label{eq:PhiR-def}
\Phi_R(\omega, \bz) \;\equiv\;
\log|\hat{\bb}-\bz| \;-\; \frac{E}{\omega}\,\log\!\left|\hat{\bb}-\frac{\omega}{E}\,\bz\right|,
\end{equation}
When the ratio $\omega/E$ is small, the second log in Eq.~\eqref{eq:PhiR-def} can be approximated to
\begin{equation}\label{eq:expansion}
\frac{E}{\omega}\,\log\!\left|\hat{\bb}-\frac{\omega}{E}\bz\right|
\;=\; -\,\hat{\bb}\cdot \bz \;+\; \mathcal{O}(\omega/E).
\end{equation}
In this approximation, the function $\Phi_R(\omega, \bz)$ simplifies to
\begin{equation}\label{eq:Phi-limit}
\Phi_R(\omega, \bz) \;\xrightarrow{\,\omega/E\,\to\, 0\,}\;
\hat{\bb}\cdot \bz \;+\; \log|\hat{\bb}-\bz|
\;\equiv\; \Phi(\bz).
\end{equation}
With this we finally arrive at the single hit matched amplitude which takes the following form:
\begin{equation}
\label{app-M1-zrep}
\mathcal{M}_{\text{matched}}(b;E,\omega,\bth)
\;\simeq\;
-s\kappa^3
\int\!\frac{d^2 \bz}{2\pi^2\bz^{*2}}\,e^{ib\omega z\cdot\bth}\,\Phi(\bz),
\end{equation}
The above expression for the gravitational emission amplitude only takes into account a single graviton exchange between the two external particles plus the emitted radiation. The full physical amplitude is obtained by convolving this result with the entire eikonal series; the Feynman diagram corresponding to the $n$'th term in the series is shown in Fig.~\ref{fig:eikonal-emission}. 
CCCV obtained this physical amplitude by averaging over the eikonal trajectory of the light particle. As the light particle traverses the H-shockwave it is deflected at the running angle $\xi\boldsymbol{\theta}_s$ where $\boldsymbol{\theta}_s=(2R/b)\hat{\bsb}$ is the total elastic deflection and $\xi\in[0,1]$ can be viewed to parameterize the position along the trajectory. The negative helicity resummed amplitude is~\cite{Ciafaloni:2015vsa, Ciafaloni:2015xsr}
\be
\label{app-resum}
\mathfrak{M}^{(-2)}(b,E;\omega,\boldsymbol{\theta}) \;=\; -s\kappa^3 \int_0^1 d\xi\int \frac{d^2 \bz}{2\pi^2\bz^{*2}}\,e^{ib\omega \bz\cdot (\boldsymbol{\theta}-\xi\boldsymbol{\theta}_s)}\,\Phi(\bz)~.
\ee
Using $b\omega\bz\cdot\boldsymbol{\theta}_s=2\omega Rx$
with $x\equiv\hat{\bsb}\cdot\bz$, the $\xi$-integral gives
\be
\label{app-sinc-emerge}
\int_0^1\!d\xi\,e^{-2i\omega Rx\,\xi} \;=\; e^{-i\omega Rx}\,\frac{\sin(\omega Rx)}{\omega Rx}\,.
\ee
Hence the final result for the negative helicity resummed amplitude is
\be
\label{app-resum-final}
\mathfrak{M}^{(-2)}(b,E;\omega,\boldsymbol{\theta}) \;=\; -s\kappa^3 \int \frac{d^2 \bz}{2\pi^2 \bz^{*2}}\, e^{ib\omega \bz\cdot\boldsymbol{\theta}} e^{-i\omega Rx}\,\frac{\sin(\omega Rx)}{\omega Rx} \,\Phi(\bz)~.
\ee
In a similar manner one can obtain the positive helicity amplitude $\mathfrak{M}^{(+2)}$. The result is
\begin{equation}
\mathfrak{M}^{(+2)}(b, E ; \omega, \theta)=-s \kappa^3 \int \frac{d^2 \bz}{2 \pi^2 \bz^2} e^{i b \omega \bz \cdot \theta} e^{+i \omega R x} \frac{\sin (\omega R x)}{\omega R x} \Phi(\bz)~.
\end{equation}
Denoting the fixed helicity resummed amplitude by $\mathfrak{M}^{(\lambda)}$, we are now in a position to calculate the total emission rate which is given by 
\begin{equation}
\label{eq:dE-int}
\frac{dE_{GW}}{d\omega}
\;=\;\omega^2\,\sum_\lambda\!
\int\!d\Omega\,\bigl|\mathfrak{M}^{(\lambda)}\bigr|^2.
\end{equation}
here $E_{GW}$ denotes the energy of the emitted gravitational wave (not to be confused with $E$, the energy of external particles). In the small angle approximation ($\sin\bth \simeq \bth $), the angular integration $\int d\Omega$ over the graviton emission direction $\bth$ becomes $\int d^2\bth$. This integral in $\bth$ can be simplified using Parseval's identity as follows. Define
\begin{equation}
F(\bz)
\equiv
\frac{1}{\bz^{*2}}\,
e^{-i\omega R x}
\frac{\sin(\omega R x)}{\omega R x}\,
\Phi(\bz),
\qquad
\boldsymbol q\equiv b\omega\,\boldsymbol\theta ~.
\end{equation}
Using $d^2\bsq=(b\omega)^2d^2\bth$ and the Parseval's identity,
\begin{equation}
\int d^2q\,
\left|
\int d^2\bz\,e^{i\boldsymbol q\cdot\bz}
F(\bz)
\right|^2
=
(2\pi)^2\int d^2\bz\,|F(\bz)|^2,
\end{equation}
the angular integral in Eq.~\eqref{eq:dE-int} for either helicity becomes
\begin{equation}
\omega^2\int d^2\bth\,
\left|\mathfrak{M}^{(\lambda)}(\boldsymbol\theta)\right|^2
=
\frac{1}{b^2}
\frac{s^2\kappa^6}{\pi^2}
\int d^2\bz\,|F(\bz)|^2 .
\end{equation}
Since,
\begin{equation}
|F(\bz)|^2
=
\frac{1}{|\bz|^4}
\left[
\frac{\sin(\omega R x)}{\omega R x}
\right]^2
|\Phi(\bz)|^2,
\end{equation}
we get the final result (the factor of 2 below comes from the sum over helicities)
\be
\label{app-master}
\frac{dE_{GW}}{d\omega} \;=\; \frac{2s^2 \kappa^6}{(\pi b)^2}\,\int \frac{d^2\bz}{|\bz|^4}\,
\left[\frac{\sin(\omega Rx)}{\omega Rx}\right]^2\,|\Phi(\bz)|^2\,.
\ee
This is one of the main results of \cite{Ciafaloni:2015vsa, Ciafaloni:2015xsr}. Notice that all of the $\omega$-dependence sits in the sine factor. 

At large $\omega R$, we can simplify this expression further. Denote
\begin{align}
a &\equiv \omega R, \\
K_a(x) &\equiv \frac{a}{\pi}
\left[\frac{\sin(ax)}{ax}\right]^2 .
\end{align}
We first note that $K_a(x)$ approaches the Dirac delta distribution as $a\to\infty$. For any smooth test function $f(x)$,
the change of variables $u=ax$ gives
\begin{align}
\int_{-\infty}^{\infty} K_a(x)f(x)\,\mathrm{d}x
&=
\frac{a}{\pi}
\int_{-\infty}^{\infty}
\left[\frac{\sin(ax)}{ax}\right]^2
f(x)\,\mathrm{d}x
\\
&=
\frac{1}{\pi}
\int_{-\infty}^{\infty}
\left(\frac{\sin u}{u}\right)^2
f\left(\frac{u}{a}\right)\,\mathrm{d}u .
\end{align}
Since
\begin{equation}
\int_{-\infty}^{\infty}
\left(\frac{\sin u}{u}\right)^2
\mathrm{d}u
=
\pi
\end{equation}
and $f(u/a)\to f(0)$ as $a\to\infty$, we have that
\begin{align}
\lim_{a\to\infty}
\int_{-\infty}^{\infty}K_a(x)f(x)\,\mathrm{d}x
&=
\frac{f(0)}{\pi}
\int_{-\infty}^{\infty}
\left(\frac{\sin u}{u}\right)^2
\mathrm{d}u =f(0)~.
\end{align}
Therefore, $K_a(x)$ satisfies the properties of Dirac delta function in the limit $a\to\infty$. Next, without loss of generality orient the $\bz$ plane such that the direction of the $x$ axis matches with the direction of impact parameter $\hat{\bsb}$. This allows us to write $\bz=(x,y)$, so that $|\bz|^2=x^2+y^2$ and
\begin{align}
\Phi(x,y)
&=
\hat{\boldsymbol{b}}\cdot\bz
+\log\left|\hat{\boldsymbol{b}}-\bz\right|
\\
&=
x+\frac{1}{2}
\log\left[(1-x)^2+y^2\right].
\end{align}
Applying the delta-function limit with $a=\omega R$ to the integral in Eq.~\eqref{app-master}, we obtain
\begin{align}
\int
\frac{\mathrm{d}^2\bz}{|\bz|^4}
\left[
\frac{\sin(\omega R x)}{\omega R x}
\right]^2
|\Phi(\bz)|^2 \xrightarrow{\omega R\gg 1}
&\frac{\pi}{\omega R}
\int_{-\infty}^{\infty}
\frac{|\Phi(0,y)|^2}{y^4}\,\mathrm{d}y~,\nonumber \\
=&\frac{\pi}{\omega R}
\left[
\frac{1}{4}
\int_{-\infty}^{\infty}
\frac{\log^2(1+y^2)}{y^4}\,\mathrm{d}y
\right]~,\nonumber\\
=&\frac{2\pi^2}{3\omega R}
\left(1-\log 2\right)~.
\end{align}

Inserting this result back in Eq.~\eqref{app-master} we recover the result 
\be
\frac{dE_{GW}}{d\omega}\bigg|_{\omega R\gg 1} \;\simeq\;  \frac{4s^2\kappa^6}{3 b^2}\,\frac{1-\log 2}{\omega R}. 
\ee

\section{Lipatov radiation equals eikonal radiation in the soft limit}
\label{app:eikonal-soft-NLO}

\begin{figure}[ht]
    \centering
    \includegraphics[scale=0.65]{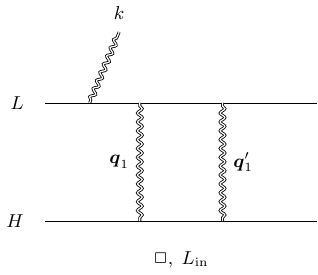}\quad
    \includegraphics[scale=0.65]{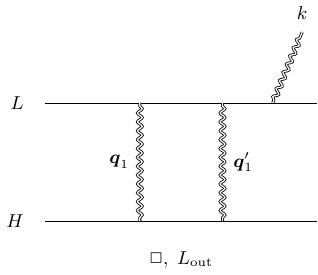}\quad
    \includegraphics[scale=0.65]{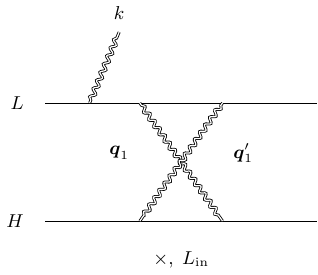}\quad
    \includegraphics[scale=0.65]{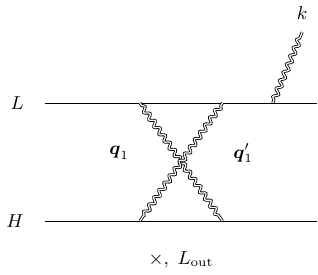}
    \caption{The four external-line diagrams that reproduce the leading soft $O(\mu_L\mu_H^2)$ contribution. $H,L$ denote the two massless particles  and the exchanged gravitons carry transverse momenta $\boldsymbol q_1$ and $\boldsymbol q_1'$.}
    \label{fig:Eik1-4}
\end{figure}

In this Appendix, we will provide an independent scattering amplitude check of the soft limit of the $O(\mu_L\mu_H^2)$ result obtained in Sec.~\ref{sec:complete-rescattering}. We consider the eikonal $2\rightarrow 2+1$ amplitude with two exchanged gravitons and attach the emitted graviton to the hard particle lines, as shown in Fig.~\ref{fig:Eik1-4}. We will explicitly demonstrate that the ``connected" contribution from these diagrams reproduce the NLO soft result in Eq.~\eqref{eq:M-soft-b-omega}. We are only interested in the leading Weinberg soft behavior $\mathcal O(\omega^{-1})$. Graviton emissions from a) external line between two $t$-channel graviton exchanges, b) from the  exchanged gravitons, and c) the gravitational self-interaction vertex on an external line, all begin at $\mathcal O(\omega^0)$ as discussed in Sec.~\ref{subsec:angular}, and therefore does not enter the discussion below. 

Another important point concerning two exchanges is that the box plus crossed-box amplitude contains the eikonal iteration of the one-exchange radiation amplitude. We will separate this iterated contribution from the connected two-graviton exchange contribution. The latter is the quantity that can be directly compared with the $O(\mu_L\mu_H^2)$ shockwave result in Eqs.~\eqref{eq:S-soft}--\eqref{eq:M-soft-b-omega}.

Let us start with the Born level (single exchange) negative helicity soft emission amplitude that takes the following form
\begin{align}
\label{eq:born-soft}
\mathcal{M}_{\text{soft}, L}^{(1),(-2)}=\mathcal{M}_{\text {el }}^{(1)} \mathcal{W}_L^{(-2)}~,
\end{align}
where $\mathcal{M}_{\text {el }}^{(1)}$ is the single graviton exchange Born amplitude
\begin{align}
\mathcal M_{\rm el}^{(1)}(\bsq) =\frac{\kappa^2s^2}{\bsq^2}~,
\label{appB:Mel-Born}
\end{align}
and $\mathcal{W}_L^{(-2)}$ is the soft emission Weinberg current attached to the light source (L), projected on to  negative helicity,
\begin{align}
\mathcal W_{L}^{(-2)}
\equiv
\frac{\kappa}{2}\left(\frac{
p_L'^\mu p_L'^\nu\epsilon_{\mu\nu}^{(-2)}
}{
p_L'\cdot k
}
-
\frac{
p_L^\mu p_L^\nu\epsilon_{\mu\nu}^{(-2)}
}{
p_L\cdot k
}\right)~.
\label{appB:WL-def}
\end{align}
With the light-cone momentum conventions
\begin{align}
p_L^\mu=\left(0, \mu_L, \boldsymbol{0}\right), \quad p_L^{\prime \mu}=\left(\frac{\mathbf{q}^2}{2 \mu_L}, \mu_L, \mathbf{q}\right),
\end{align}
the eikonal propagators are determined to be
\begin{align}
    p_L\cdot k = \mu_L k^+ = \frac{\bsk^2}{2x_L}~,\qquad
    p_L'\cdot k = \frac{1}{2x_L}|\tilde{k}-x_L\tilde q|^2
\end{align}
where we defined $x_L=\frac{k^-}{\mu_L}$.
Next, with $\epsilon^{(-)}$ defined in Eq.~\eqref{polarization-tensor}, we have
\begin{equation}
p_L \cdot \epsilon^{(-)}=\frac{\tilde{k}^*}{\sqrt{2} x_L}~,\qquad p_L^{\prime} \cdot \epsilon^{(-)}=\frac{\tilde{k}^*-x_L \tilde{q}^*}{\sqrt{2} x_L} .
\end{equation}
Using the above, we then evaluate Eq.~\eqref{appB:WL-def} to be
\begin{align}
\mathcal W_{L}^{(-2)}(\boldsymbol q,\boldsymbol k)
=
\frac{\kappa}{2}e^{-2i\phi_k}
\frac{
\tilde k^*\tilde q-\tilde k\tilde q^*
}{
\tilde k^*
\left(
\tilde k-x_L\tilde q
\right)
}~.
\label{appB:WL-exact}
\end{align}
In the high energy limit, $x_L$ is small and $\mathcal W_{L}^{(-2)}$ admits the Taylor expansion,
\begin{align}
\mathcal{W}_L^{(-2)}=\mathcal{W}_{L, \mathrm{LE}}^{(-2)}+\mathcal{W}_{L, \mathrm{NE}}^{(-2)}+\cdots
\end{align}
where the leading eikonal $\mathcal{W}_{L,\mathrm{LE}}^{(-2)}$ and the next-to-leading eikonal $\mathcal{W}_{L,\mathrm{NE}}^{(-2)}$ terms are
\begin{align}
\mathcal{W}_{L,\mathrm{LE}}^{(-2)}=\frac{\kappa}{2}e^{-2 i \phi_k} \frac{\tilde{k}^* \tilde{q}-\tilde{k} \tilde{q}^*}{\bsk^2}\,\,\,;\,\,\,
\mathcal{W}_{L,\mathrm{NE}}^{(-2)}=\frac{\kappa}{2}\frac{k^{-}}{\mu_L} e^{-2 i \phi_k} \frac{\left(\tilde{k}^* \tilde{q}-\tilde{k} \tilde{q}^*\right) \tilde{q}}{\bsk^2 \tilde{k}}~.
\label{eq:W-NE}
\end{align}
Decomposing the transverse momentum with respect to the direction of $\bsk$,
$\bsq
=
q_\parallel\hat{\boldsymbol k}
+
q_\perp\hat{\boldsymbol n}$
and 
$\hat{\boldsymbol n}\cdot\hat{\boldsymbol k}=0$, the expression for $\mathcal{W}_{L, \mathrm{NE}}$ becomes
\begin{align}
\mathcal W_{L,{\rm NE}}^{(-2)}(\bsq,\bsk)
=
-\frac{\kappa k^-}{\mu_L\bsk^2}\,
e^{-2i\phi_k}
\left(
q_\perp^2
-iq_\parallel q_\perp
\right)~.
\label{appB:WNE-real}
\end{align}
The corresponding term from the heavy ($H$) source is obtained by
$L\leftrightarrow H$ and $+\leftrightarrow-$ and generates the complementary $O(\mu_L^2\mu_H)$ contribution. Since the dilute-dense expansion studied in the main text is restricted to first order in $\mu_L$, our comparison below will retain only the $L$-line contribution.

We now consider eikonal scattering of two massless particles $H$ and $L$ at $O(\kappa^4)$ where two gravitons with transverse momenta $\bsq_1$ and $\bsq_1'$ are exchanged with a net momentum transfer $\boldsymbol Q = \bsq_1+\bsq_1'$. For each of the box and crossed-box topologies, the emitted graviton can be attached either before the first exchange or after the second exchange on the hard line. (The emission from between the two exchanges is subleading in the soft regime). As noted, the relevant diagrams that contribute to the shockwave result Eq.~\eqref{eq:M-soft-b-omega} are shown in Figs.~\ref{fig:Eik1-4}.

These four diagrams in the regime of soft graviton emission can be written as
\begin{align}
\sum_{a={\rm in,out}}
\mathcal M_{\Box,L,a}^{(-2)}
&=
\mathcal W_L^{(-2)}(\boldsymbol Q,\bsk)
\mathcal M_{\Box}^{(2)}
+\mathcal O(\omega^0)~,
\label{appB:box-L}
\\[3pt]
\sum_{a={\rm in,out}}
\mathcal M_{\times,L,a}^{(-2)}
&=
\mathcal W_L^{(-2)}(\boldsymbol Q,\bsk)
\mathcal M_{\times}^{(2)}
+\mathcal O(\omega^0)~,
\label{appB:cross-L}
\end{align}
where $\mathcal M_{\Box}^{(2)}$ and $\mathcal M_{\times}^{(2)}$ are the {\it elastic} amplitudes corresponding to the double exchange box and cross-box diagrams.

The sum of the double exchange box and crossed-box elastic amplitudes in the leading eikonal approximation is the two-fold convolution of the single exchange Born amplitude \cite{Ciafaloni:2014esa}
\begin{align}
\mathcal M_{\rm el}^{(2)}(\boldsymbol Q)
&\equiv\mathcal M_{\Box}^{(2)}(\boldsymbol Q)
+
\mathcal M_{\times}^{(2)}(\boldsymbol Q)
=
\frac{i}{4s}
\int\frac{d^2\bsq_1}{(2\pi)^2}\,
\mathcal M_{\rm el}^{(1)}(\bsq_1)
\mathcal M_{\rm el}^{(1)}
(\boldsymbol Q-\bsq_1)
\nonumber\\
&=
\frac{i\kappa^4s^3}{4}
\int\frac{d^2\bsq_1}{(2\pi)^2}
\frac{1}{
\bsq_1^2\bsq_1'^2
},
\qquad
\bsq_1'
\equiv
\boldsymbol Q-\bsq_1 .
\label{appB:second-Born}
\end{align}
Using Eqs.~\eqref{appB:box-L} and \eqref{appB:cross-L}, the full
$L$-line two-exchange soft amplitude in impact-parameter space is
therefore
\begin{align}
\mathcal M_{{\rm full},L}^{(2),(-2)}(\bsb;k)
=
\frac{i\kappa^4s^3}{4}
\int\frac{d^2\bsq_1}{(2\pi)^2}
\frac{d^2\bsq_1'}{(2\pi)^2}
\frac{
e^{i(\bsq_1+\bsq_1')\cdot\bsb}
}{
\bsq_1^2\bsq_1'^2
}
\mathcal W_L^{(-2)}
(\bsq_1+\bsq_1',\bsk).
\label{appB:Mraw}
\end{align}

Eq.~\eqref{appB:Mraw} contains an eikonal iteration in which the one-graviton exchange soft radiation amplitude is accompanied by an additional elastic exchange. This ``disconnected"  contribution is
\begin{align}
\mathcal M_{{\rm disconn.},L}^{(2),(-2)}(\bsb;k)
={}&
\frac{i}{4s}
\int\frac{d^2\bsq_1}{(2\pi)^2}
\frac{d^2\bsq_1'}{(2\pi)^2}
e^{i(\bsq_1+\bsq_1')\cdot\bsb}
\big[
\mathcal M_{{\rm soft},L}^{(1),(-2)}(\bsq_1;k)
\mathcal M_{\rm el}^{(1)}(\bsq_1')
+
\bsq_1\leftrightarrow \bsq_1'
\big]
\no\\[10pt]
=&
\frac{i\kappa^4s^3}{4}
\int\frac{d^2\bsq_1}{(2\pi)^2}
\frac{d^2\bsq_1'}{(2\pi)^2}
\frac{
e^{i(\bsq_1+\bsq_1')\cdot\bsb}
}{
\bsq_1^2\bsq_1'^2
}
\left[
\mathcal W_L^{(-2)}(\bsq_1,\bsk)
+
\mathcal W_L^{(-2)}(\bsq_1',\bsk)
\right]\,,
\label{appB:Miter}
\end{align}
where $\mathcal{M}_{{\rm soft},L}^{(1),(-2)}$ and $\mathcal{M}_{\rm el}^{(1)}$ were defined in Eqs.~\eqref{eq:born-soft} and \eqref{appB:Mel-Born} respectively. The second term in the bracket denotes an identical product to the first term with $\bsq_1\leftrightarrow \bsq_1'$.
The connected two-graviton exchange
contribution is therefore 
\begin{align}
\mathcal M_{{\rm conn.},L}^{(2),(-2)}
\equiv
\mathcal M_{{\rm full},L}^{(2),(-2)}
-
\mathcal M_{{\rm disconn.},L}^{(2),(-2)}\,.
\label{appB:Mconn-def}
\end{align}
Using Eqs.~\eqref{appB:Mraw} and \eqref{appB:Miter},
\begin{align}
\mathcal M_{{\rm conn},L}^{(2),(-2)}(\bsb;k)
=
\frac{i\kappa^4s^3}{4}
\int\frac{d^2\bsq_1}{(2\pi)^2}
\frac{d^2\bsq_1'}{(2\pi)^2}
\frac{
e^{i(\bsq_1+\bsq_1')\cdot\bsb}
}{
\bsq_1^2\bsq_1'^2
}
\Delta\mathcal W_L^{(-2)},
\label{appB:Mconn}
\end{align}
where
\begin{align}
\Delta\mathcal W_L^{(-2)}
\equiv{}&
\mathcal W_L^{(-2)}
(\bsq_1+\bsq_1',\bsk)
-
\mathcal W_L^{(-2)}
(\bsq_1,\bsk)
-
\mathcal W_L^{(-2)}
(\bsq_1',\bsk).
\label{appB:DeltaW}
\end{align}

The leading high-energy Weinberg current in Eq.~\eqref{eq:W-NE} is linear in the transverse momentum transfer and therefore satisfies
\begin{align}
\mathcal W_{L,{\rm LE}}^{(-2)}
(\bsq_1+\bsq_1',\bsk)
=
\mathcal W_{L,{\rm LE}}^{(-2)}
(\bsq_1,\bsk)
+
\mathcal W_{L,{\rm LE}}^{(-2)}
(\bsq_1',\bsk).
\end{align}
Hence, $\Delta\mathcal W_{L,{\rm LE}}^{(-2)}=0$; the leading eikonal part of the full two-graviton exchange soft amplitude is entirely contained in the disconnected term in 
Eq.~\eqref{appB:Miter}. The first nonvanishing connected contribution is therefore  the first subleading high energy term, which can be represented as in Eq.~\eqref{appB:WNE-real}. Applying Eq.~\eqref{appB:DeltaW} to Eq.~\eqref{appB:WNE-real}, with
\begin{align}
\bsq_1
=
q_{1\parallel}\hat{\boldsymbol k}
+
q_{1\perp}\hat{\boldsymbol n}~,~\qquad
\bsq_1'
=
q_{1'\parallel}\hat{\boldsymbol k}
+
q_{1'\perp}\hat{\boldsymbol n},
\end{align}
gives
\begin{align}
\Delta\mathcal W_{L,{\rm NE}}^{(-2)}
=
-\frac{2k^-}{\mu_L\bsk^2}
e^{-2i\phi_k}
\Big[
2q_{1\perp}q_{1'\perp}
-i\left(
q_{1\parallel}q_{1'\perp}
+
q_{1'\parallel}q_{1\perp}
\right)
\Big]~.
\label{appB:DeltaWNE}
\end{align}
The terms in the square brackets are precisely 
$\mathcal N$ introduced below Eq.~\eqref{eq:M-soft-before-integrals}:
\begin{align}
\mathcal N
=
2q_{1\perp}q_{1'\perp}
-i\left(
q_{1\parallel}q_{1'\perp}
+
q_{1'\parallel}q_{1\perp}
\right).
\label{appB:N}
\end{align}
Further, Eq.~\eqref{appB:DeltaWNE} has exactly the same
transverse structure as the soft limit of
$\widetilde S^{(-2)}$ in Eq.~\eqref{eq:S-soft}. Substituting Eq.~\eqref{appB:DeltaWNE} into Eq.~\eqref{appB:Mconn} gives
\begin{align}
\mathcal M_{{\rm conn.},L}^{(2),(-2)}(\bsb;k)
=&
\frac{
2\mu_L\mu_H^2\kappa^5(-ik^-)
}{
\bsk^2
}
e^{-2i\phi_k}
\int\frac{d^2\bsq_1}{(2\pi)^2}
\frac{e^{i\bsq_1\cdot\bsb}}{\bsq_1^2}
\int\frac{d^2\bsq_1'}{(2\pi)^2}
\frac{e^{i\bsq_1'\cdot\bsb}}{\bsq_1'^2}
\mathcal N ~,
\label{appB:exact-momentum-match}
\end{align}
where we used $s=2\mu_L\mu_H$. 
This result is identical to Eq.~\eqref{eq:M-soft-factorized} obtained directly from the $O(\mu_L\mu_H^2)$ shockwave calculation. The remaining transverse Fourier transforms have already been evaluated
in Sec.~\ref{sec:Lipatov-soft}. From Eq.~\eqref{eq:soft-double-FT},
\begin{align}
\int\frac{d^2\bsq_1}{(2\pi)^2}
\frac{e^{i\bsq_1\cdot\bsb}}{\bsq_1^2}
\int\frac{d^2\bsq_1'}{(2\pi)^2}
\frac{e^{i\bsq_1'\cdot\bsb}}{\bsq_1'^2}
\mathcal N
=
\frac{i}{2\pi^2b^2}
e^{i\Delta\phi}
\sin\Delta\phi
=
\frac{1}{4\pi^2b^2}
\left(
e^{2i\Delta\phi}-1
\right),
\end{align}
where $\Delta\phi=\phi_b-\phi_k$. The rest follows identically to the discussion in Sec.~\ref{sec:Lipatov-soft}, and we obtain
\begin{align}
\mathcal M_{{\rm conn},L}^{(2),(-2)}(\bsb;k)
=
-\frac{
i\mu_L\mu_H^2\kappa^5
}{
4\sqrt2\pi^2b^2
}
\frac{1}{\omega}
e^{-2i\phi_k}
\left(
e^{2i\Delta\phi}-1
\right)\,,
\label{appB:final-omega}
\end{align}
which is Eq.~\eqref{eq:M-soft-b-omega} in the main text. Thus the soft pole obtained in the dilute-dense shockwave calculation to $O(\mu_L\mu_H^2)$ is reproduced by the connected part of the two-graviton exchange eikonal Feynman diagrams, when the soft graviton emission is from the light source.

\bibliographystyle{JHEP.bst}
\addcontentsline{toc}{section}{References}
\bibliography{references}

\end{document}